\documentclass[10pt,preprint]{aastex}

\usepackage{mathtools}
\usepackage{amsmath}
\usepackage{relsize}
\usepackage{booktabs}
\usepackage{comment}
\usepackage{natbib}
\usepackage{graphicx}
\usepackage[caption=false]{subfig}
\usepackage{enumitem}
\usepackage{relsize}

\newcounter{eqn}

\makeatletter
\newcommand{\putindeepbox}[2][0.7\baselineskip]{{%
    \setbox0=\hbox{#2}%
    \setbox0=\vbox{\noindent\hsize=\wd0\unhbox0}
    \@tempdima=\dp0
    \advance\@tempdima by \ht0
    \advance\@tempdima by -#1\relax
    \dp0=\@tempdima
    \ht0=#1\relax
    \box0
}}
\makeatother

\usepackage[usenames,dvipsnames]{xcolor}

\usepackage{color}

\shorttitle{Modeling radio foregrounds for EoR}
\shortauthors{Sathyanaraynara Rao et al.}

\begin{document}

\title{GMOSS: All-sky model of spectral radio brightness based on physical components and associated radiative processes}

\author{Mayuri Sathyanarayana Rao$^{1,2}$, Ravi Subrahmanyan$^{1}$, N Udaya Shankar$^{1}$, Jens Chluba$^{3}$}
\affil{{\small $^{1}$Raman Research Institute, C V Raman Avenue, Sadashivanagar, Bangalore 560080, India}}
\affil{{\small $^{2}$Australian National University, Research School for Astronomy \& Astrophysics, Mount Stromlo Observatory, Cotter Road, Weston, ACT 2611, Australia}}
\affil{{\small $^{3}$Jodrell Bank Centre for Astrophysics, University of Manchester, Oxford Road, M13 9PL, U.K.}}
\email{Email of corresponding author: mayuris@rri.res.in}

\begin{abstract}
We present Global MOdel for the radio Sky Spectrum (GMOSS) -- a novel, physically motivated model of the low-frequency radio sky from 22 MHz to 23 GHz. GMOSS invokes different physical components and associated radiative processes to describe the sky spectrum over 3072 pixels of $5^{\circ}$ resolution.  The spectra are allowed to be convex, concave or of more complex form with contributions from synchrotron emission, thermal emission and free-free absorption included. Physical parameters that describe the model are optimized to best fit four all-sky maps at 150 MHz, 408 MHz, 1420 MHz and 23 GHz and two maps at 22 MHz and 45 MHz generated using the Global Sky Model of \citet[]{GSM2008}. The fractional deviation of model to data has a median value of $6\%$ and is less than $17\%$ for $99\%$ of the pixels. Though aimed at modeling of foregrounds for the global signal arising from the redshifted 21-cm line of Hydrogen during Cosmic Dawn and Epoch of Reionization (EoR) - over redshifts $150\lesssim z \lesssim 6$, GMOSS is well suited for any application that requires simulating spectra of the low-frequency radio sky as would be observed by the beam of any instrument. The complexity in spectral structure that naturally arises from the underlying physics of the model provides a useful expectation for departures from smoothness in EoR foreground spectra and hence may guide the development of algorithms for EoR signal detection. This aspect is further explored in a subsequent paper.
\end{abstract}

\section{Introduction}
Interest in the low frequency radio sky goes back to the very early days of radio astronomy. Early measurements at $20.5$ MHz by Karl Jansky showed that the Galaxy itself is a strong emitter of radiation at low frequencies. The source of such Galactic emission was poorly understood and was attributed to a wide variety of phenomena right from radio stars to dust grains. It is now understood that the predominant radiative mechanism contributing to this Galactic emission at long wavelengths is synchrotron radiation from relativistic electrons spiraling around Galactic magnetic field lines. With many experiments currently underway attempting to detect redshifted 21-cm spectral signatures arising from Cosmic Dawn and the Epoch of Reionization, there is renewed interest in understanding the precise spectral shapes of the low frequency radio sky, which forms a strong foreground to the weak cosmological signal. 

The radio sky is composed of signals from our Galaxy, extragalactic radio sources and the cosmic microwave background (CMB), with added spectral distortions related to the cosmic thermal history of baryons, structure formation, energy release in the early Universe and interactions between propagating radiation and gas. These include distortions from Cosmic Dawn and the Epoch of Reionization. The emergence of the first sources in the Cosmic Dawn and the transition of baryonic matter in the Universe from being almost completely neutral to its present mostly ionized form during the epoch of reionization (EoR) is an interesting and poorly constrained period in Cosmology. During these times the cosmological evolution of the spin temperature of Hydrogen and reionization as a result of first light from the first collapsed objects is expected to leave an imprint as redshifted 21 cm emission and absorption \citep[see][]{Madau1997,Shaver1999}. These 21-cm spectral distortions are a probe of the thermal history of the gas and also the sources and timing of reionization \citep{Glover2014} in the redshift range 6 to about 150. For a comprehensive review of the subject refer to \cite{Furlanetto2006}. There are global, all-sky isotropic spectral features as well as angular variations in spectral structure, embedded as tiny additive components in the radio background at frequencies $\lesssim 200~$MHz. Radio emission from Galactic and extragalactic sources forms strong foregrounds to the cosmological signal and are orders of magnitude brighter.  

It is necessary to have a realistic expectation for the radio foreground that would be observed by EoR detection experiments \citep[see][]{Bowman2008,Patra2013,Voytek2014,Bernardi2015,Pober2014,Pober2015,Ali2015}. Although the sky spectrum as measured by individual experiments will be instrument-specific, a generic model representative of the spectral distribution of intensities in the low-frequency radio sky and a method to simulate the expected contribution of the same to spectra observed in EoR detection telescopes will be a powerful tool in formulating data analysis methods.  

Here we present a physically-motivated model of the low-frequency radio-sky: Global MOdel for the radio Sky Spectrum (GMOSS). GMOSS is a generic model and can be used to generate spectra of the low-frequency radio sky for other applications as well. One such application is to simulate the expected foregrounds for other spectral distortions of the CMB such as those arising from the Epoch of Recombination \citep{Sunyaev2009}. As noted in \cite{apsera2015}, the optimal frequency for a ground based detection of cosmological recombination lines is an octave band in the range of 2--6 GHz. Over these frequencies the recombination signal is expected to have a quasi-periodic sinusoidal shape, whereas the foregrounds are expected to be smooth. However, since the cosmological recombination signal is at least eight orders of magnitude weaker than the foreground, a thorough treatment of the expectation of the spectral shapes inherent in the foreground can be provided by GMOSS.   

The motivation for GMOSS in comparison to existing sky-models is given in Section \ref{sec:motivation}. GMOSS itself is described in Section~\ref{sec:skymodel} and a discussion on the distribution of parameters and goodness of fit is presented in Section~\ref{sec:results}.

\section{Motivation}
\label{sec:motivation}

Precise measurement of foreground spectra is a goal of experiments aiming to detect redshifted 21-cm signatures from Cosmic Dawn and the Epoch of Reionization (henceforth referred to in totality as the EoR). Disentangling spectral structure specific to the EoR signal from those of the foreground \citep{Harker2015} and instrument \citep{Switzer2014} is a challenging problem. In generic models \citep[such as Figure 1 of][]{Pritchard2010b}, the redshifted 21-cm signal (in the 10 to 200~MHz window) is expected to show multiple turning points  arising from various physical processes such as X-ray and UV heating, ionization of the gas and Wouthuysen-Field coupling of spin to kinetic temperature via Lyman-$\alpha$ to name a few \citep[for a description of the physics that determines the turning points in the EoR signal, see][]{Pritchard2010a}. On the other hand, it is assumed that over the same frequency range foregrounds are smooth \citep[see][]{Petrovic2011} and that `smoothness' is captured by low-order polynomials in log brightness temperature versus log frequency space (we hereinafter refer to this domain as log-log space). It has been proposed to exploit this supposed smoothness of foregrounds to distinguish them from the global EoR signal, which in the generic form is expected to show multiple turning points between $10$ and $200$ MHz. However, it is uncertain if this assumption of an inherently smooth foreground is indeed true. For instance, a mechanism that might result in inflections of the observed spectrum is the combined emission from steep and flat spectrum sources along with radiation from sources that have a break in the electron energy distribution. Furthermore, flattening of this combined spectrum due to absorption at low frequencies by thermal  interstellar medium along the line of sight and at high frequencies by free-free emission can introduce additional shapes in the spectrum. Though the assumption is that averaging of multiple spectra of various shapes across the sky and along the line of sight must result in an observed spectrum that is devoid of sharp spectral features, the underlying spectral energy distribution (SED) that results in the radiation would ultimately determine the inherent smoothness of the observed spectrum. An interpolating function that is guided by physical processes provides a non-trivial treatment of the shape of the foreground spectrum. 

\cite{Waelkens2009} present a tool to simulate maps of the total and polarised synchrotron emission of the radio sky including effects of Faraday rotation. Other popular sky-models such as the Global Sky Model (GSM;  \citet[]{GSM2008} with recent improvements made by \citet[]{Zheng2016}) use data-driven methods to generate snapshots of the sky at frequencies between those where large area maps are available. This is useful in creating all-sky maps at discrete frequencies and may also be used to generate spectra in any sky direction by computing the sky brightness over a contiguous range of frequencies. It is estimated that for sub-GHz frequencies predicted maps from GSM are at worst in error by $\lesssim~10\%$ depending on the region of the sky. To date in the literature mock spectra have been generated by first computing the beam-weighted temperatures from maps at discrete frequencies and then interpolating with either power laws or low-order polynomials to simulate EoR foregrounds.  For instance, \cite{Pritchard2010a} generate a mock spectrum by allowing for GSM generated sky to drift over the zenith of an ideal frequency-independent cos$^2$ antenna beam for 24 hours. They find that their mock spectra can be fit to mK precision with a cubic polynomial in log-log space. Independently, \cite{Bernardi2015} find that to fit the spectrum of synchrotron emission from mono-energetic cosmic ray electrons to within $100~$mK maximum errors, a polynomial of sixth order is required in log-log space. For the case of synchrotron emission from an evolved population of cosmic ray electrons diffusing through the Galactic halo, the order of the polynomial required reduces to four. \cite{Harker2016} produce mock sky spectra that contain foregrounds generated using polynomial interpolation and fit them with polynomials of similar order to remove the foregrounds, eliminating any ambiguity arising from using polynomials of a different order. While this helps in focusing on statistical inference methods for the parameters of the cosmological signal itself, it remains to be examined what the precise shape of the foreground spectrum might be and the extent to which it might confuse detection of the 21-cm signal from EoR. 

What is required is a model of the foreground, which encapsulates the underlying physics that gives rise to the foreground spectrum. Whether or not the resultant spectrum is smooth would guide the strategies employed to detect the EoR signal that is embedded in a measurement set. If using polynomials to model the foreground spectrum, the order of the polynomial required to fit the foreground with the precision required for detection of the EoR signal should be guided by the spectral complexity determined the underlying physical radiative processes.  

In GMOSS we present a physical sky model that includes effects of plausible radiative processes arising from different components of the radio sky. Additionally, physical parameters used to describe GMOSS are guided by all-sky maps at different frequencies, thus enabling a check on the goodness of the model.

\section{Global MOdel for the radio Sky Spectrum (GMOSS) : A spatially-resolved spectral model for the radio sky }
\label{sec:skymodel}

Previous efforts to determine the complexity of EoR foregrounds have usually simulated sky spectra by interpolating with polynomials between brightness temperatures that were first computed for large instrument beam widths \citep[see][]{Harker2015}. However, even if spectra of individual sources that lie in the beam are of power-law form, the summation of many such spectra with a distribution in spectral indices across the beam and along the line of sight would result in an observed spectrum that would no longer be a power law but spectrally more complex. Thus there exists a need to generate a generic sky-model at higher resolutions than typical single antenna beams, which may then be convolved with a large antenna beam to generate mock spectra that are qualitatively more representative of the cumulative emission. GMOSS provides such a sky-model wherein plausible physical processes are used to estimate the sky-spectrum toward each direction. Furthermore the parameters describing the spectral shape towards each sky pixel and hence describing the sky-model are constrained by existing all-sky maps. 

We use 4 available all-sky maps at 150~MHz \citep{Landecker1970}, 408~MHz \citep{Haslam1982}, 1420~MHz \citep{Reich1982, Reich1986, Gorski2005} and  23 GHz (WMAP science data product \footnote{WMAP Science Team.}) to generate GMOSS. Additionally, we also use all-sky maps at 22~MHz and 45~MHz generated using the Global Sky Model \citep{GSM2008}. These last two all-sky maps are expected to closely match the raw data maps that were inputs to the GSM at the corresponding frequencies. The CMB monopole temperature was subtracted from the maps, when present.  All the maps were reduced to a common resolution of $5^{\circ}$ and represented with the `R4' nested HEALPix\footnote{\url{http://healpix.sourceforge.net}} scheme in galactic coordinates. Additionally, the temperature scale of the 150~MHz all-sky map of \citet{Landecker1970} was corrected by subtracting an offset of 21.4~K and scaling the pixel intensities by a factor 1.05 to improve the accuracy of the representation \citep[see][]{Patra2015}. An offset of $493~\mu$K was added to the 23 GHz map to include an estimate of the uniform component missing in the differential map, as described in \cite{apsera2015}. Using the resultant images, with identical beams and pixelation, we generate a physical sky model for the Galactic and Extragalactic emission (excluding the CMB) in the 22~MHz--23~GHz band, as described below.

\subsection{GMOSS: Physics}
\label{sec:physics}

\begin{figure}[htp]
\begin{minipage}{0.5\textwidth}
\subfloat[]{%
  \includegraphics[clip,width=\columnwidth]{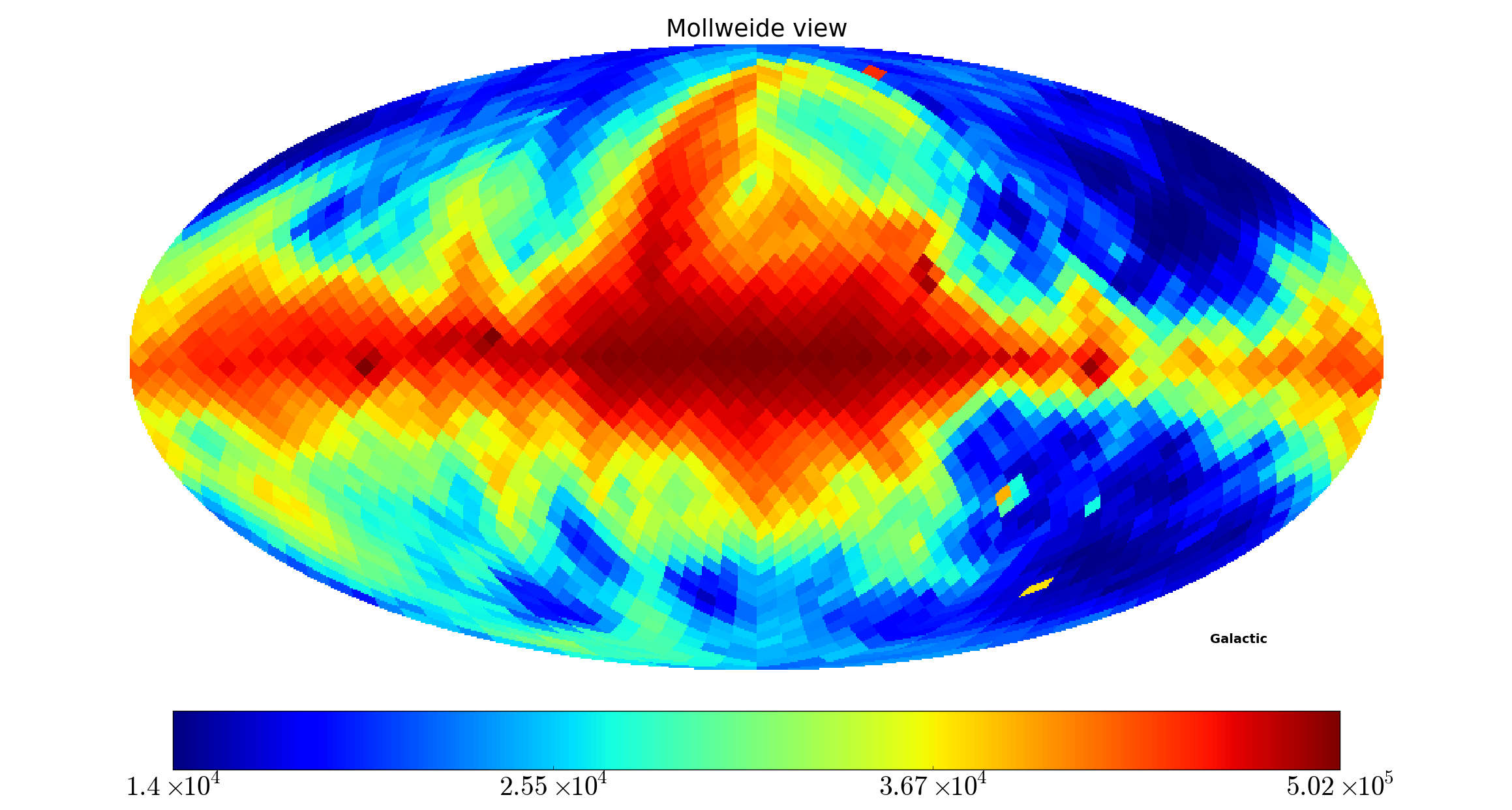}%
}

\subfloat[]{%
  \includegraphics[clip,width=\columnwidth]{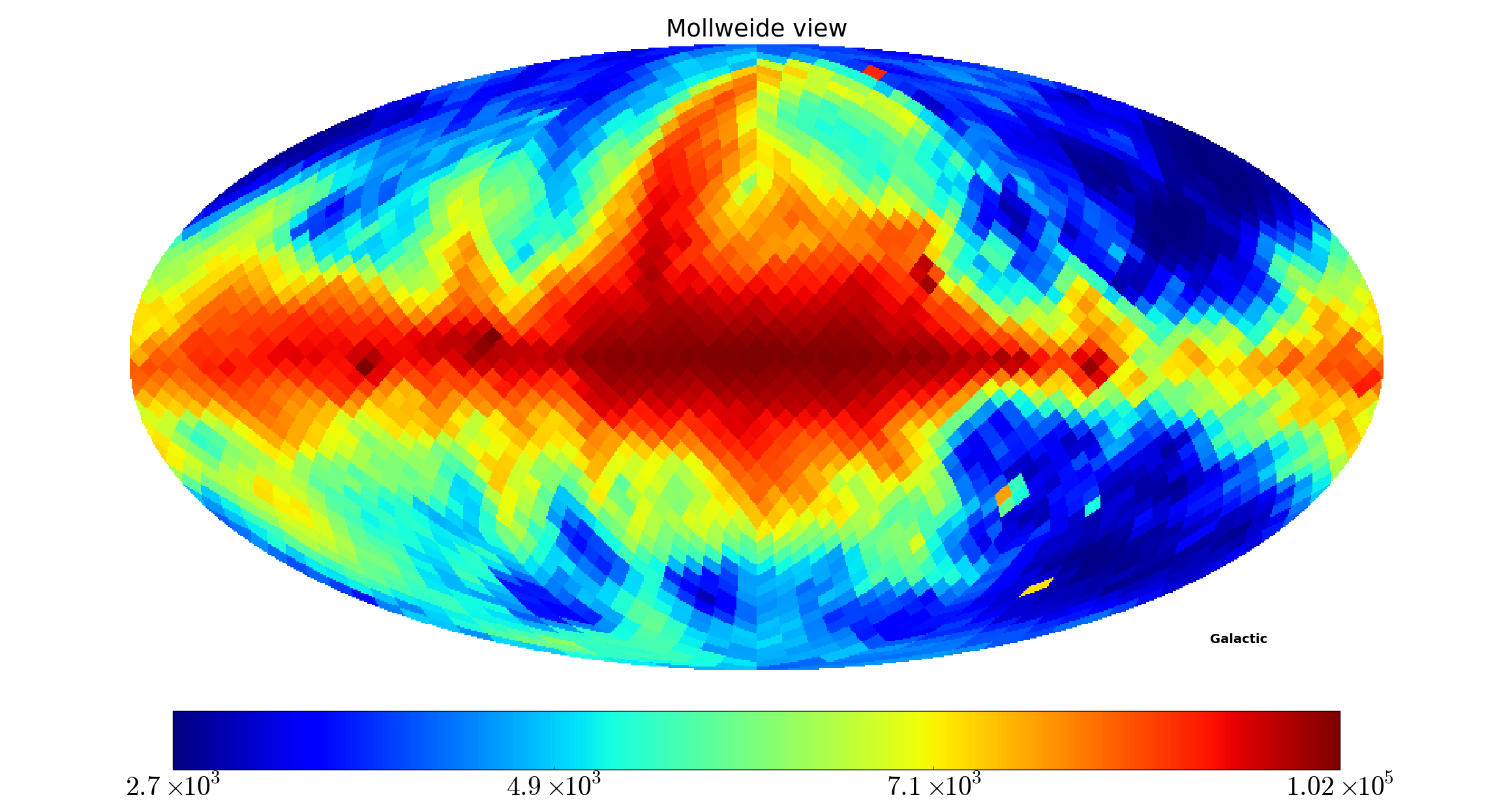}%
}

\subfloat[]{%
  \includegraphics[clip,width=\columnwidth]{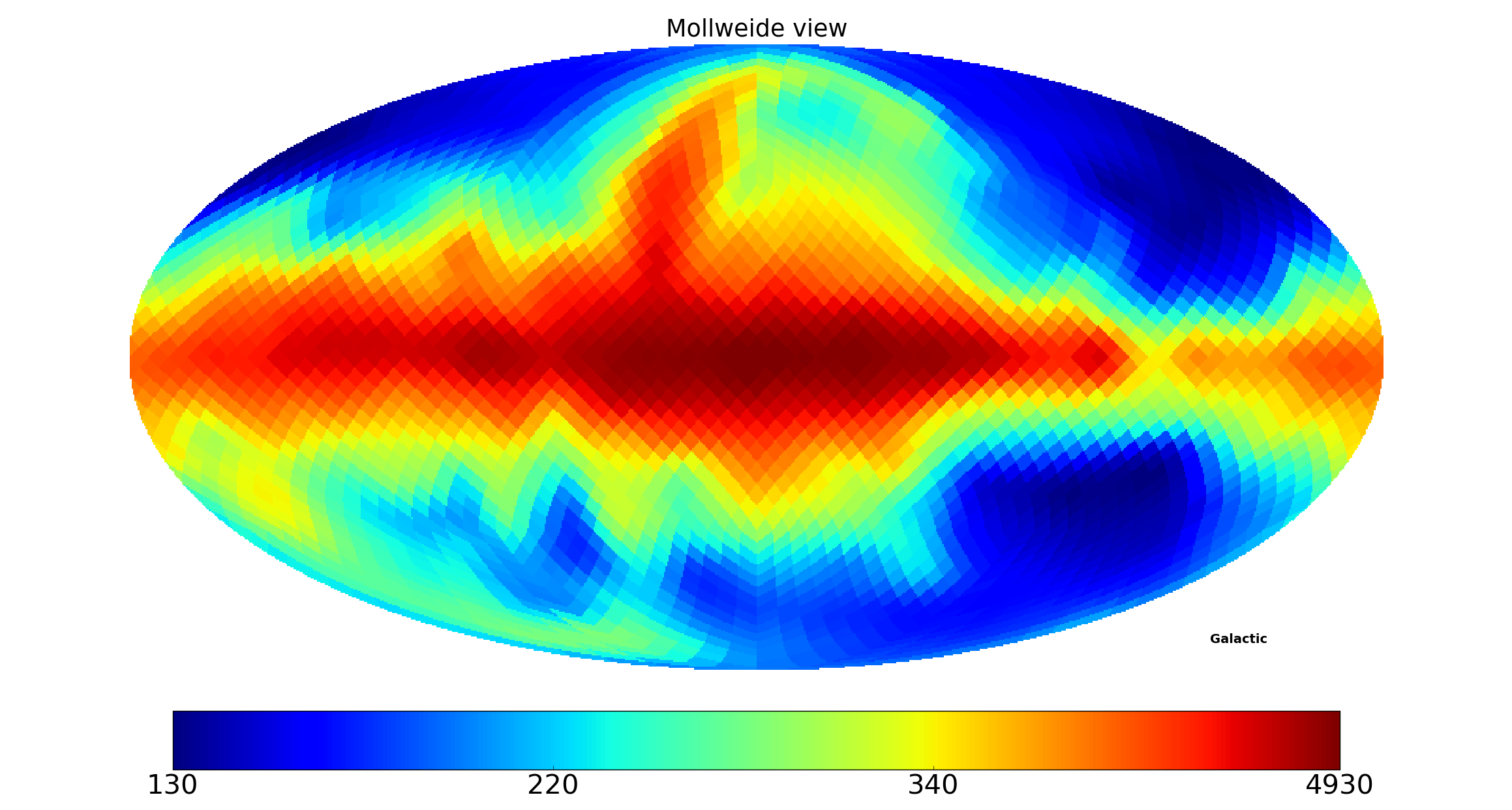}%
}
\end{minipage}
\begin{minipage}{0.5\textwidth}

\subfloat[]{%
  \includegraphics[clip,width=\columnwidth]{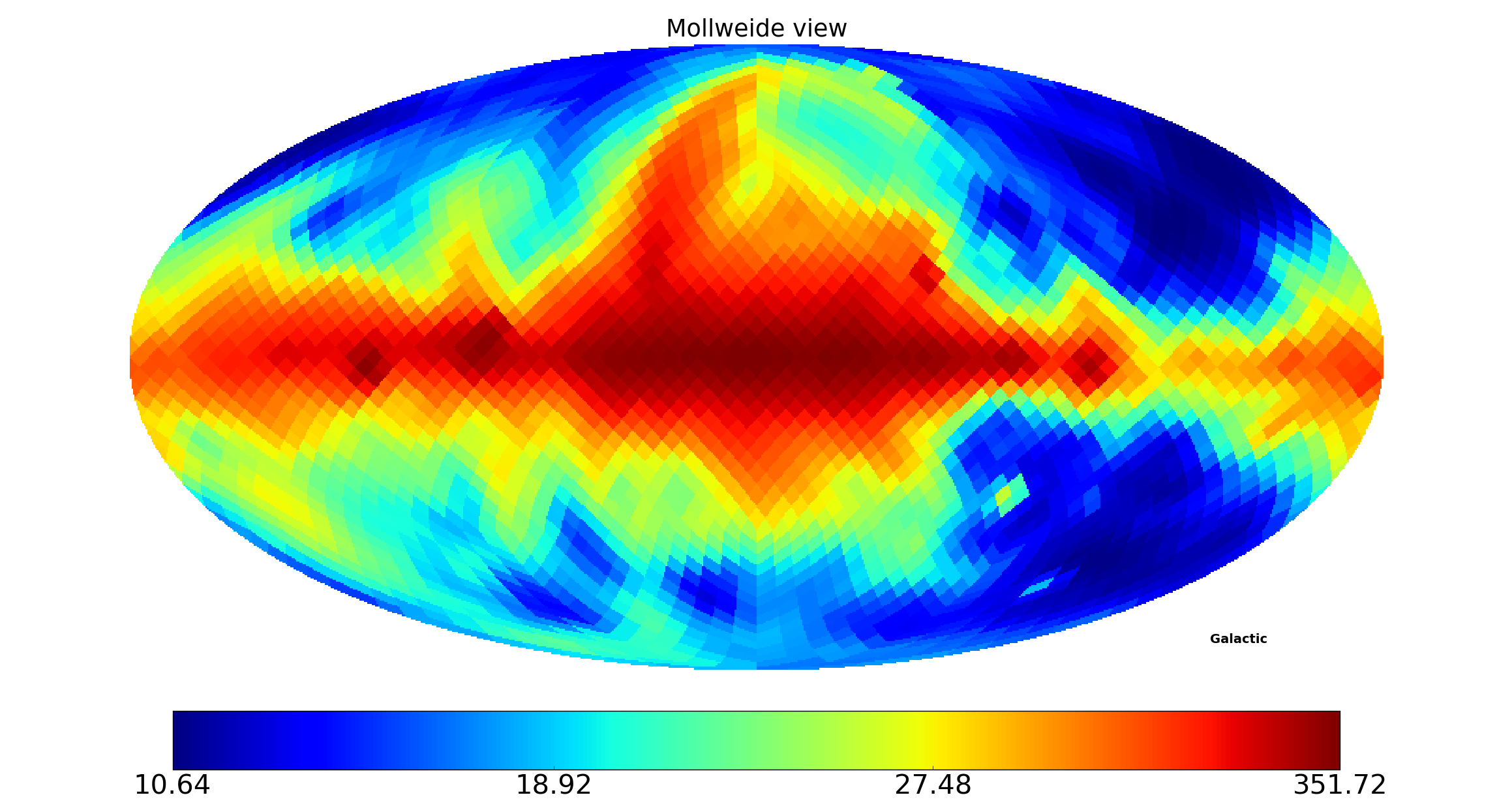}%
}

\subfloat[]{%
  \includegraphics[clip,width=\columnwidth]{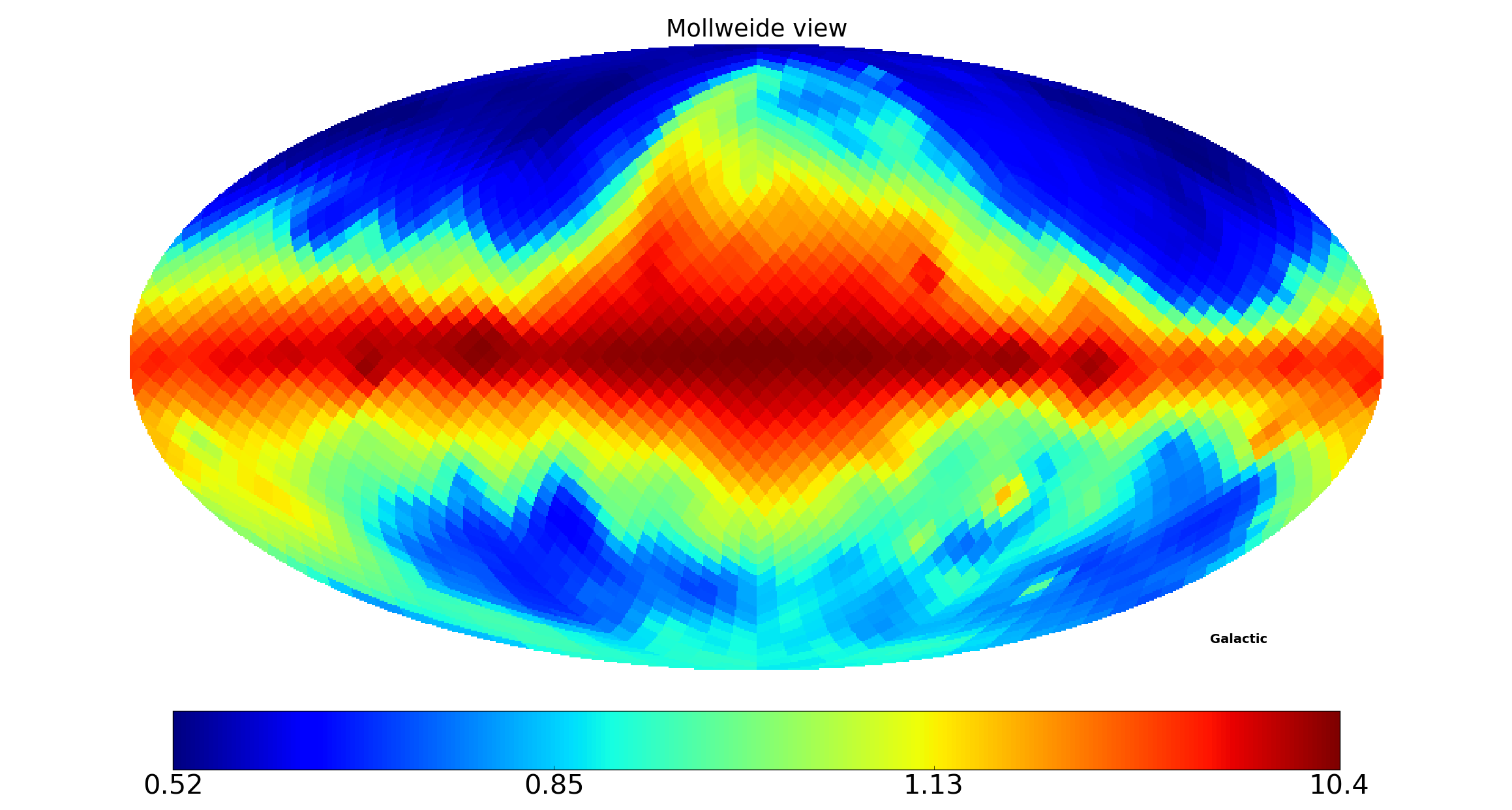}%
}

\subfloat[]{%
  \includegraphics[clip,width=\columnwidth]{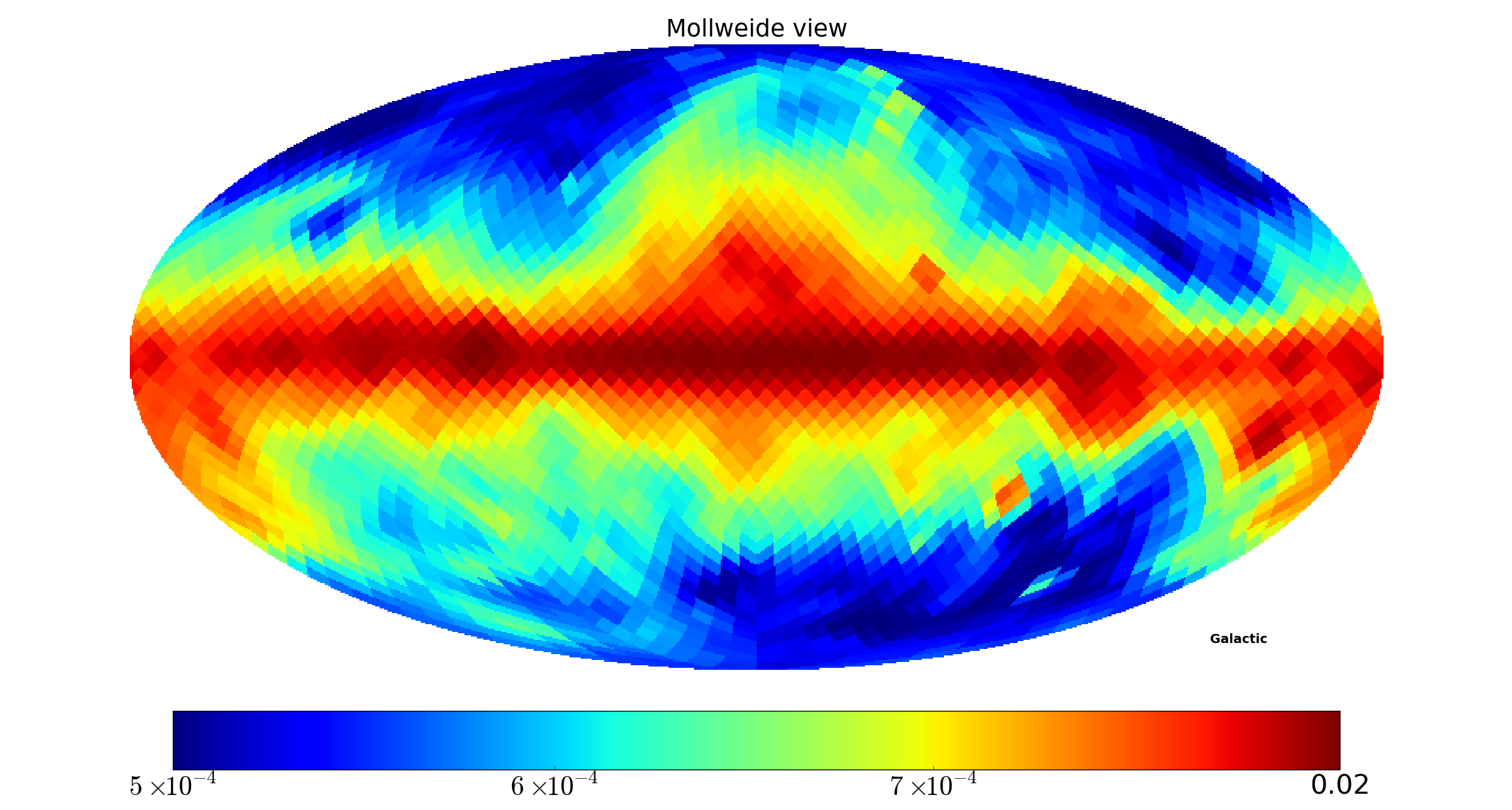}%
}
\end{minipage}
\caption{Maps used as input to GMOSS. Maps at (a) 22 MHz and (b) 45 MHz are generated from GSM \citep{GSM2008}. Map (c) is at 150~MHz \citep{Landecker1970} with corrections applied from \citet{Patra2015}. Map (d) 408~MHz \citep{Haslam1982}, (e) 1420~MHz \citep{Reich1982, Reich1986, Gorski2005} and (f)  23 GHz (WMAP science data product$^1$) with a uniform component added as described in \cite{apsera2015}.
All maps are in units of Kelvin at a common resolution of $5^{\circ}$ and in galactic coordinates, represented in with nested `R4' scheme of HEALPix.}\label{fig:inp_maps}
\end{figure}

Herein we describe GMOSS. The six maps used as inputs to generate GMOSS are shown in Figure~\ref{fig:inp_maps}. 

The dominant mechanism of emission at low radio frequencies is synchrotron radiation. The total spectrum emitted by an ensemble of electrons $N(\gamma)$, with a distribution $N(\gamma)d\gamma\propto\gamma^{\rm -p}d\gamma$ and energies ranging between $\gamma_{\rm min}$ and $\gamma_{\rm max}$, is given by
\begin{equation}
P_{\rm tot}(\nu) = C\int\limits_{\gamma_{\rm min}}^{\gamma_{\rm max}} P(\nu)\gamma^{\rm -p} d\gamma	.\label{eq:Ptot}\\
\end{equation}

Here $\gamma$ is the Lorentz factor of the electrons, indicative of electron energy and $P(\nu)$ is the emission spectrum from individual electrons. $p$ is the index of electron energy distribution $N(\gamma)$. We define the temperature spectral index `$\alpha$' such that brightness temperature $T(\nu)~\propto~\nu^{\rm -\alpha}$, wherein $\alpha$ is related to the electron distribution index by $\alpha = \frac{\rm p+3}{2}$.
Further, $P(\nu) \propto F(x)$ with
\begin{equation}
F(x) =  x\int\limits_{x}^{\infty} K_{\rm \frac{5}{3}}(\xi) d\xi ,	\label{eq:Fofx}\\
\end{equation}
where $x = \frac{\nu}{\nu_{\rm c}}$ and $K_{\rm \frac{5}{3}}$ is the Bessel function of the second kind. $\nu_{\rm c}$ is the critical frequency given by
\begin{align}\label{eq:nuc}
\nu_{\rm c} &= \frac{3\gamma^2qB\textrm{sin}\alpha}{4\pi~mc},
\end{align}
where $B$ is the magnetic field, $\alpha$ the pitch angle between particle (electron) velocity and magnetic field, $c$ the velocity of light in free space, $q$ the charge of the particle (electron) and $m$ its mass. We refer the reader to \cite{rl1986} for a detailed treatment of the synchrotron radiation process. 

\begin{figure}[htp]
\begin{minipage}{0.5\textwidth}

\subfloat[Concave]{%
  \includegraphics[clip,width=\columnwidth, height = 2.0in]{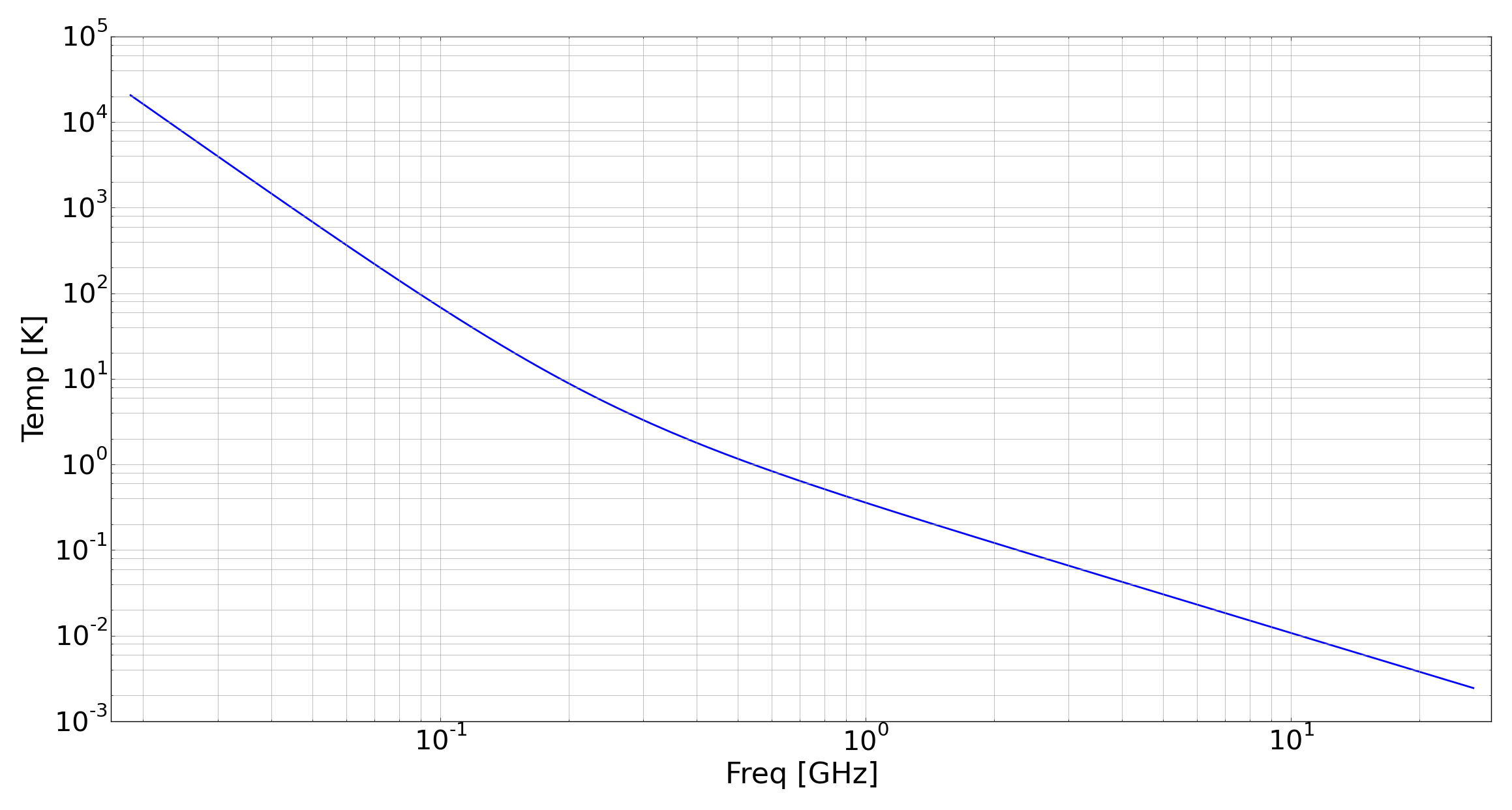}%
}
\end{minipage}
\begin{minipage}{0.5\textwidth}

\subfloat[Convex]{%
  \includegraphics[clip,width=\columnwidth, height = 2.0in]{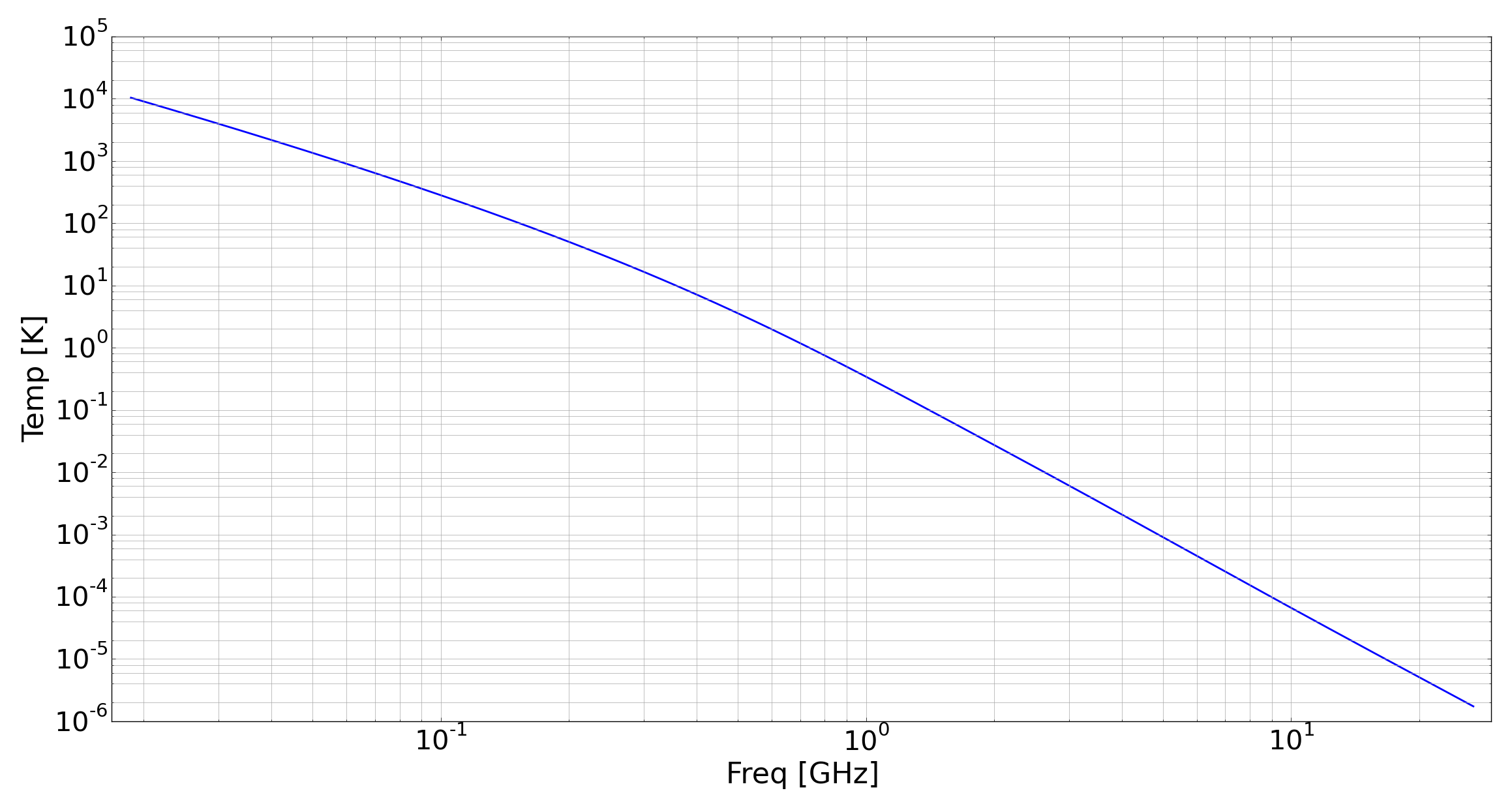}%
}
\end{minipage}

\begin{minipage}{0.5\textwidth}

\subfloat[More complex shape that is predominantly concave]{%
  \includegraphics[clip,width=\columnwidth, height = 2.0in]{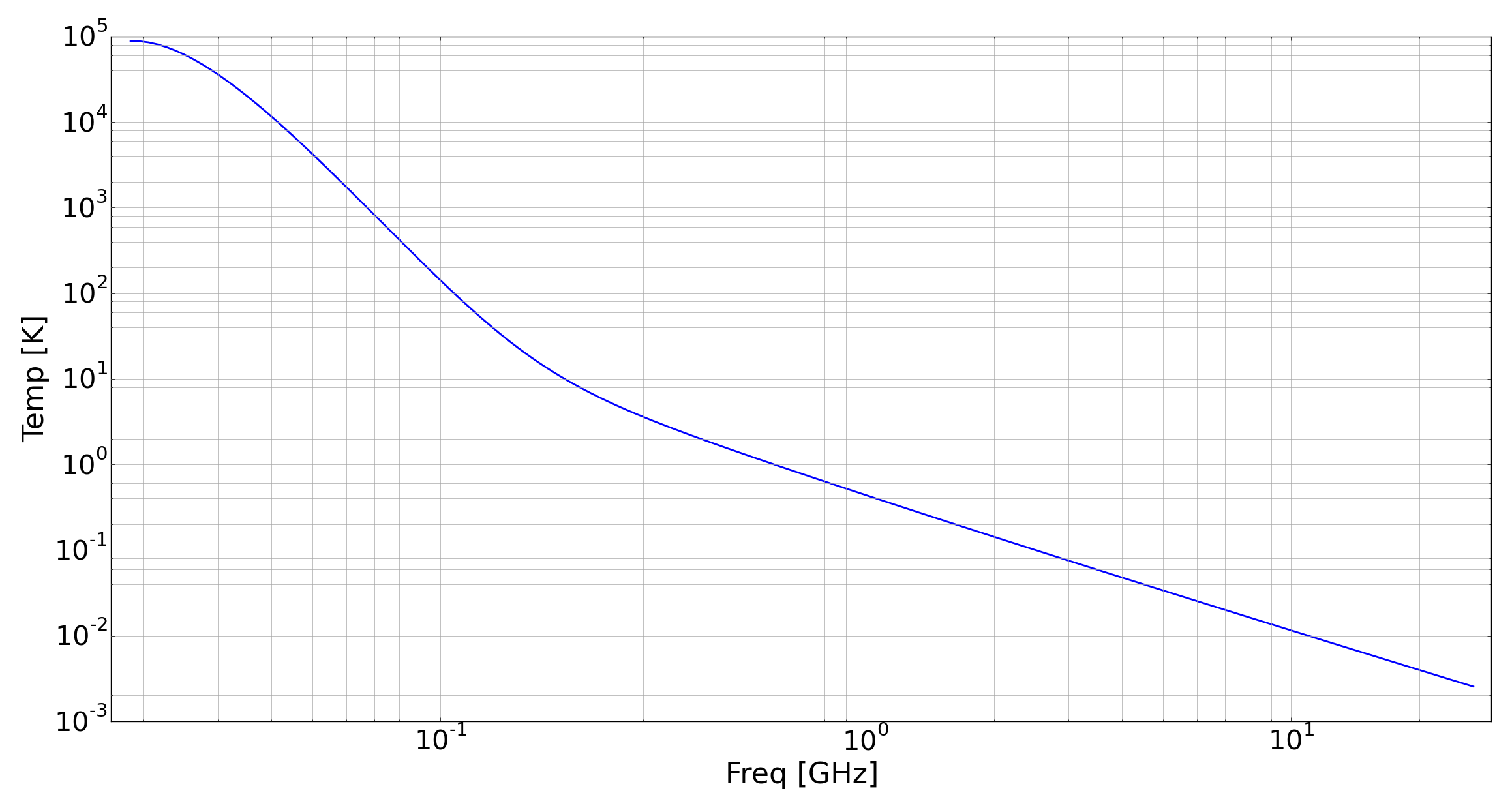}%
}
\end{minipage}
\begin{minipage}{0.5\textwidth}

\subfloat[More complex shape that is predominantly convex]{%
  \includegraphics[clip,width=\columnwidth, height = 2.0in]{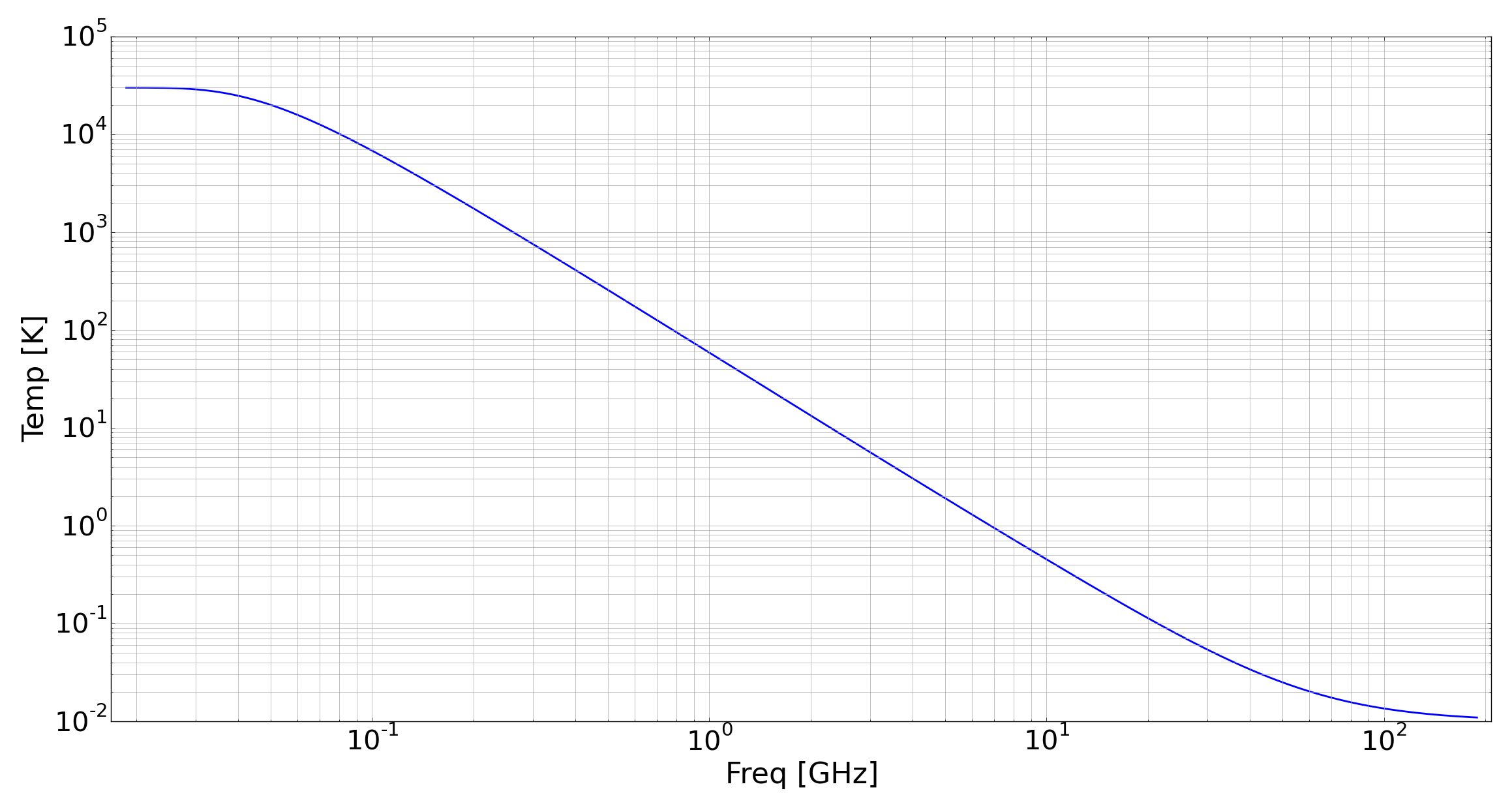}%
}
\end{minipage}
\caption{Toy-models showing exaggerated spectral shapes in log-temperature versus log-frequency scale. Panel (a) shows a concave spectrum which can arise from a combination of flat and steep spectral sources, the steep component dominates at low frequencies and the flat component at high frequencies. The curve in (b) traces a spectrum which has a convex shape arising form a break in the electron energy distribution.  Examples of more complex spectral shapes are shown in (c) and (d). While the curve in (c) shows a concave spectrum that is flattened at low frequencies by thermal absorption, the curve in (d) shows the case of a convex spectrum that is flattened by thermal absorption at low frequencies and free-free emission at high frequencies. Note that these figures are to be considered only for purposes of representation of spectral shapes.}\label{fig:toy_model}
\end{figure}

Spectral shapes may be described as convex, concave or more complex. Toy models of each of the shapes, exaggerated for purposes of representation, are shown in Figure~\ref{fig:toy_model}. In GMOSS, convex spectra are modeled as synchrotron emission arising from an ensemble of electrons having a break in their energy distribution, with steeper energy indices at higher frequencies. Concave spectra are modeled as composites of steep and flat spectrum components wherein the steep spectral component dominates at lower frequencies and flat spectral component at higher frequencies. Spectra with more complex shapes are modeled as convex or concave with significant additional thermal absorption at low frequencies and/or optically thin free-free emission at high frequencies.    
The parameters that describe the physical model optimally are $C_{\rm 1}$, $\alpha_{\rm 1}$, $\delta_{\rm \alpha}$, $\nu_{\rm br}$, $C_{\rm 2}$, $T_{\rm e}$, $I_{\rm x}$ and $\nu_{\rm t}$.  We denote by $\alpha_{\rm 1}$ and $\alpha_{\rm 2}$ the low and high frequency spectral indices respectively for the synchrotron component of the model;  parameter $\delta_\alpha$ is defined by the relation: $\alpha_{\rm 2} = \alpha_{\rm 1}+\delta_{\rm \alpha}$. The remaining parameters are described below where we separately consider the modeling of pixels with synchrotron components that are convex and concave. 

\begin{itemize}[leftmargin=*]
\item Case of pixels where $\alpha_{\rm 2} > \alpha_{\rm 1}$: We refer to such spectra as convex. We model such spectra as primarily arising from synchrotron emission from electrons with power-law energy distributions with a break, which consequently causes the emission spectrum to be a broken power law and hence of convex form. We also allow for thermal absorption at low frequencies and optically-thin thermal emission towards higher frequencies, to account respectively for any low frequency flattening and for any high frequency excess. The functional form of the model describing the sky brightness temperature $T(\nu)$ is given by 
\begin{multline}
\label{eq:break} 
T(\nu) = C_{\rm 1}\Bigg(\nu^{-2}\Big\{\gamma_{\rm break}^{2\alpha_{\rm 1} -3}\int\limits_{\gamma_{\rm min}}^{\gamma_{\rm break}} F(x)~x~\gamma^{-(2\alpha_{\rm 1} -3)}\,d\gamma~+~\gamma_{\rm break}^{2\alpha_{\rm 2} -3}\int\limits_{\gamma_{\rm break}}^{\gamma_{\rm max}} F(x)~x~\gamma^{-(2\alpha_{\rm 2} -3)}\,d\gamma\Big\}~+ \\
I_{\rm x}\nu^{-2.1}\Bigg)~e^{-(\frac{\nu_{\rm t}}{\nu})^{2.1}} + T_{\rm e}\Bigg(1 - e^{-(\frac{\nu_{\rm t}}{\nu})^{2.1}}\Bigg).
\end{multline}
The two spectral indices $\alpha_{\rm 1}$ and $\alpha_{\rm 2} = \alpha_{\rm 1}+\delta_{\rm \alpha}$ ( $\delta_{\rm \alpha} > 0$) are appropriately converted to electron distribution indices to derive the emission spectra.  The break frequency $\nu_{\rm br}$ is related to the Lorentz factor $\gamma_{\rm break}$ at which the energy spectrum has a break in its power-law form; the two are related by Equation~(\ref{eq:nuc}).  An optically-thin thermal component with brightness parameterized by $I_{\rm x}$ is added to the synchrotron emission, to account for any excess at high frequencies.  The total emission is assumed to be absorbed at a turn-over frequency $\nu_{\rm t}$ by a separate thermal foreground medium; this medium is assumed to have an electron temperature $T_{\rm e}$ and constant emission measure and would also add its own emission to the sky brightness. This is captured by the terms ${\rm e}^{\rm -(\nu_t/\nu)^{2.1}}$ and $T_{\rm e}\Bigg(1 - {\rm e}^{-(\frac{\nu_{\rm t}}{\nu})^{2.1}}\Bigg)$ respectively.  Finally the normalization parameter $C_{\rm 1}$ provides the scaling to match the observational data in temperature units. 

\item Case of pixels where $\alpha_{\rm 2} < \alpha_{\rm 1}$ : Spectra of these pixels are concave and are modeled as a composite of flat and steep spectrum synchrotron emission components with individual brightness temperatures of spectral indices $\alpha_{\rm 1}$ and $\alpha_{\rm 2} = \alpha_{\rm 1}+\delta_{\rm \alpha}$ ($\delta_{\rm \alpha} < 0$). Concave spectra could, in principle, be modeled as arising from concave energy distributions; however, that would be unphysical and hence not meaningful.  Additionally, it is also computationally easier to model concave spectra as composed of steep and flat spectra components instead of modeling as concave electron energy distribution. The functional form adopted here is given by 
\begin{equation}
\label{eq:therm}
\begin{split}
\MoveEqLeft
T(\nu) = C_1\Bigg(\nu^{-\alpha_{\rm 1}} + \frac{C_{\rm 2}}{C_{\rm 1}}\nu^{-\alpha_{\rm 2}} + I_{\rm x}~\nu^{-2.1}\Bigg){\rm e}^{-(\frac{\nu_{\rm t}}{\nu})^{2.1}} + T_{\rm e}\Bigg(1 - e^{-(\frac{\nu_{\rm t}}{\nu})^{2.1}}\Bigg).
\end{split}
\end{equation}
We drop the parameter $\nu_{\rm br}$ and instead have a second normalization parameter $C_{\rm 2}/C_{\rm 1}$, which denotes the ratio of contributions from flat and steep spectrum components. The other parameters are the same as in the case of convex spectra. In both cases there are seven parameters to be fitted for.
\end{itemize}

\subsection{GMOSS - Methods}

We employ the Downhill Simplex algorithm \citep{Nelder1965} to optimize the seven free physical-parameters of convex and concave spectral models in GMOSS by minimizing a goodness of fit $\chi^2$ towards every pixel.  $\chi^2$ is computed to be:
\begin{equation}\label{eq:chisq}
\chi^{\rm 2}=\frac{1}{N}\mathlarger{\mathlarger{\sum}\limits_{i=1}^{N}}\Bigg(\frac{T_{\rm data}(\nu_{\rm i})-T_{\rm model}(\nu_{\rm i})}{T_{\rm data}(\nu_{\rm i})}\Bigg)^2,
\end{equation}
where $T_{\rm data}$ denotes the image data at frequency $\nu_{\rm i}$, $T_{\rm model}$ denotes the model prediction at the frequency corresponding to the image data, and $N$ is the number of data points available at the sky pixel at which the fit is being done.
This constrains the model to fit the measurements with minimal fractional errors, which is appropriate if all the radio images have the same fractional errors. Despite each sky pixel being modeled with a larger number of parameters (7) than data points (6), the model spectrum for any pixel does not exactly fit all the data points for that pixel.  The deviations from the data are owing to the defined nature of the model in terms of physically motivated constraints and emission processes that allow only a specific family of curves. The high dimensionality of the modeling problem requires considerable sophistication in algorithm to obtain a good fit that converges to the global minimum in the multi-dimensional parameter space. Markov Chain Monte Carlo (MCMC) is more computationally expensive compared to the Downhill Simplex algorithm adopted herein. Future upgrades of GMOSS will adopt an implementation using MCMC as more all-sky radio images become available. The model can accommodate more input maps, including maps with partial sky coverage and can also weight images appropriate to their individual fractional errors. 

Initial guess for parameters $\alpha_{\rm 1}$ and $\alpha_{\rm 2}$ (and hence $\delta_{\rm \alpha}$) are obtained by computing 2-point temperature spectral indices between respectively the data at 45 and 150~MHz and between 408 and 1420~MHz. These determine the model employed to be either the convex or concave form.  Initial guess for normalization parameters $C_{\rm 1}$ and $C_{\rm 2}$  are evaluated at 1420~MHz. Initial guess for parameters $T_{\rm e}$ and $\nu_{\rm t}$ are set to the nominal expectations of 8000~K and 1~MHz respectively.  $I_{\rm x}$ is computed from the difference between the brightness temperature in the data at 23~GHz and that corresponding to the model evaluated after omitting the optically-thin high-frequency emission; however, if the difference is negative, then the initial guess for this parameter is set to be vanishingly small.   Further, to aid the optimization towards realistic solutions, certain parameters are constrained to be within physically acceptable limits. The temperature spectral indices are constrained to lie between 2.0 and 3.0 \citep[][]{Bennett1992} and the physical temperature $T_{\rm e}$ for the thermal foreground component that provides the low-frequency thermal¯ absorption is not allowed to exceed 10,000~K \citep[][]{Haffner2009}.
 
The synchrotron spectrum requires integration over spectra arising from individual electrons, where a single electron spectrum itself requires integration of the modified Bessel function of the second kind. Choice of the numerical integration technique is critical to accurately and efficiently implementing GMOSS. While an adaptive integration method aids in hastening computing time, care must be taken so that the numerical approximations in the algorithm do not introduce unphysical discontinuities or excessive error in the spectrum owing to computational noise. We use a combination of adaptive integration methods to optimize speed of computation and numerical accuracy. Adaptive Gauss-Kronrod quadrature is used for all integrals while estimating the parameters and the open Rhomberg adaptive method is used for integrals that generate the output spectra in specific bands where an accurate representation is desired.  For example, for generating foreground spectra that may serve for evaluating algorithms for detecting EoR signatures, the open Rhomberg adaptive method is used for integrals that generate the spectrum in a contiguous set of frequencies in the 40--200~MHz band of interest to EoR science. The method of integration allows for trade off between speed or accuracy as desired. 

The bottleneck in computing time is significantly reduced by using analytic approximations for the integral of the Bessel function. A first order analytical approximation for the integral is given in \cite{rl1986}. This is further simplified using the mathematical software tool `Mathematica' \citep{math} to reduce the number of integrations by an order of magnitude. For values of $x\ge$3 the Equation~(\ref{eq:Fofx}) can be approximated by: 
\begin{multline}
F(x) =  \frac{1}{967458816~\sqrt{2}~x^{\rm \frac{5}{2}}}~{\rm e}^{\rm -x}\Bigg(13~\sqrt{\pi} ~\Bigg\{2429625 + 2x\Big(-1922325 + 5418382x + 83221732~x^{\rm 2}\Big)\Bigg\} \\ - 1196306216~{\rm e}^{\rm x}~\pi~x^{\rm 7/2} \rm Erfc\Big[\sqrt{x}\Big] \Bigg) .	 	\label{eq:fx_large}\\
\end{multline}
Here Erfc is the complementary error function. 
For $x< 3$, Equation~(\ref{eq:Fofx}) is approximated by:
\begin{multline}
F(x) =  -\frac{\pi~x}{\sqrt{3}} + \frac{9~x^{\rm 11/3}~\Gamma\Big[-\frac{2}{3}\Big]}{160~2^{\rm 2/3}} - \frac{x^{\rm \frac{1}{3}}~\Big(16~+~3~x^{\rm2}\Big)~\Gamma\Big[-\frac{1}{3}\Big]}{24~2^{\rm 1/3}} .
	\label{eq:fx_small}\\
\end{multline}
Here, $\Gamma(x)$ is the Gamma function. Both approximations deviate from the exact treatment by less than $0.1\%$.

\section{GMOSS parameter values and goodness of fit}\label{sec:results}

\begin{figure}[htp]
\centering
\subfloat[]{%
  \includegraphics[clip,width=\columnwidth]{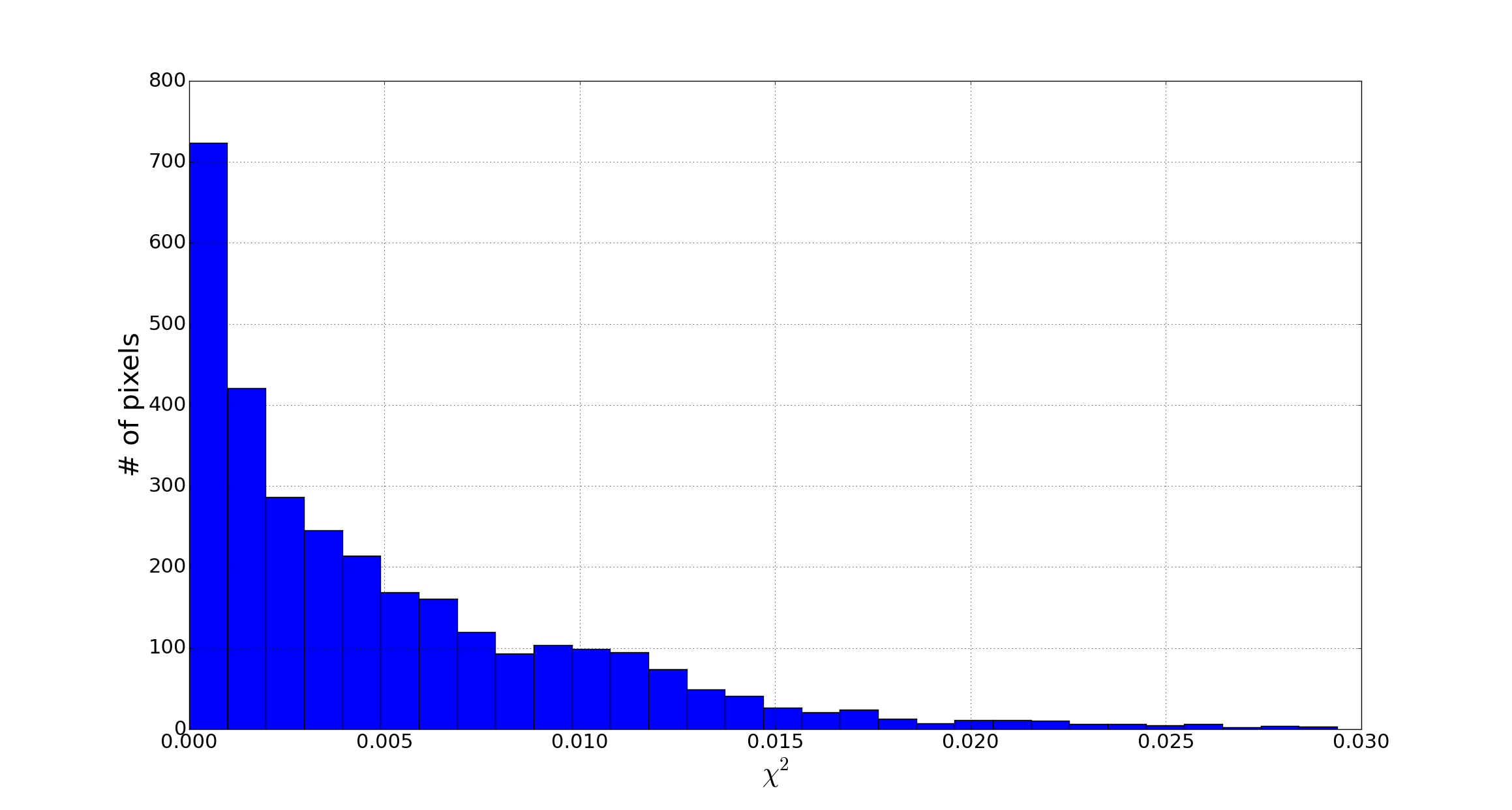}%
}

\subfloat[]{%
  \includegraphics[clip,width=\columnwidth]{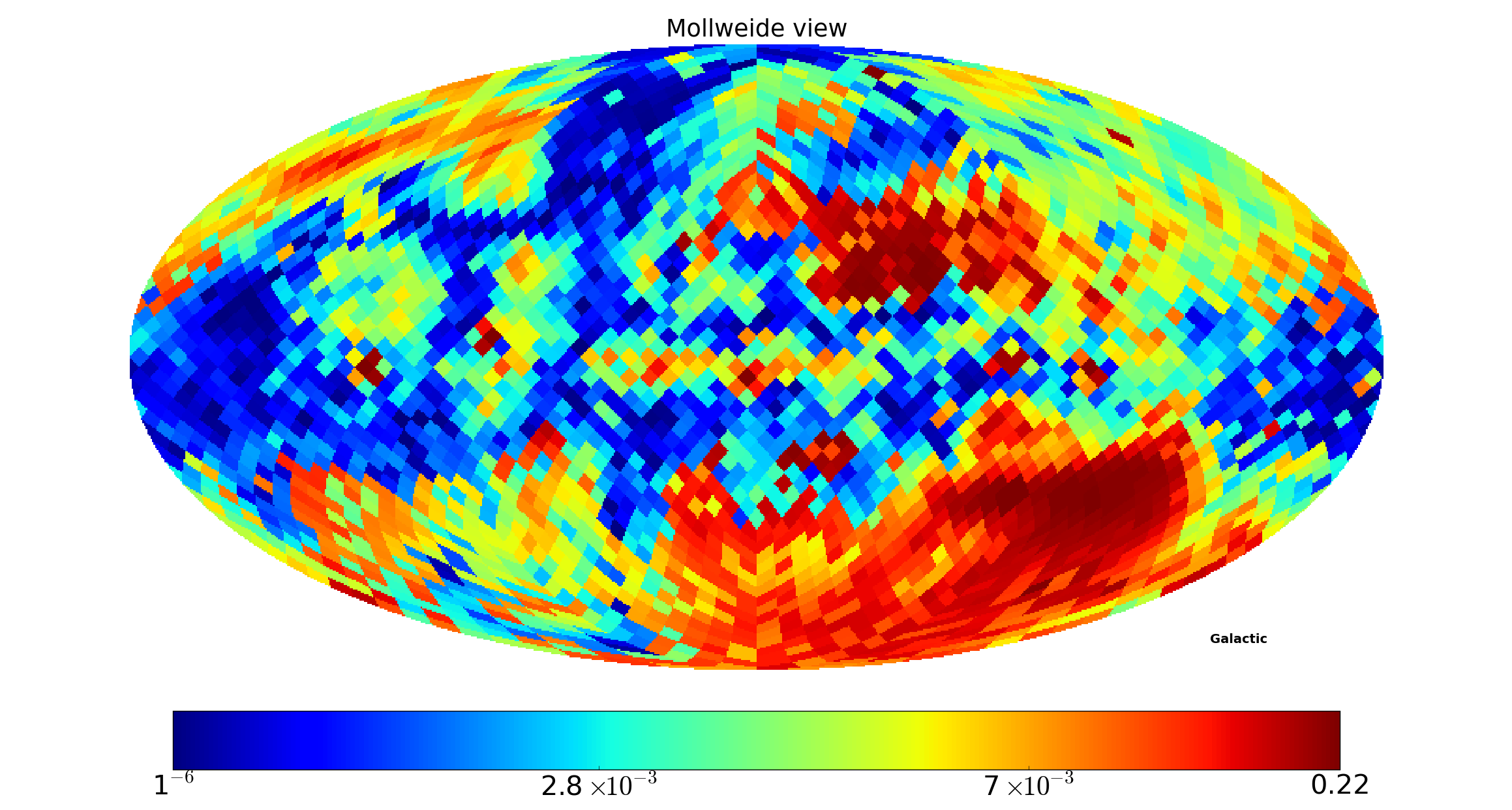}%
}
\caption{(a) Histogram of goodness of fit $\chi^2$ computed for all 3072 pixels. The median of the distribution is 0.0034, mean is 0.0054 and maximum value is 0.2280. (b) The distribution of the $\chi^2$ values computed over the 3072 sky pixels is shown in Mollweide projection, Galactic coordinates with $5^{\circ}$ resolution. }
\label{fig:chisq_99}
\end{figure}

The physical sky model fits the data at all 3072 pixels with a median $\chi^2$ of 0.0034, corresponding to a mean fractional departure of $6\%$.  $99\%$ of the pixels have  $\chi^2$ less than 0.03, corresponding to mean fractional error of $17\%$, and a histogram of $\chi^2$ is shown in Figure~\ref{fig:chisq_99}.   Also shown in Figure \ref{fig:chisq_99} is an all-sky map of $\chi^2$ on a Mollweide projection of the sky in Galactic coordinates.  
Surprisingly, the errors are relatively lower where the Galaxy dominates, both in the plane and in the region of the north polar spur.  In these regions the model corresponding to a convex spectrum was selected by the fitting algorithm.  

\begin{figure}[htp]
\begin{minipage}{0.5\textwidth}

\subfloat[]{%
  \includegraphics[clip,width=\columnwidth]{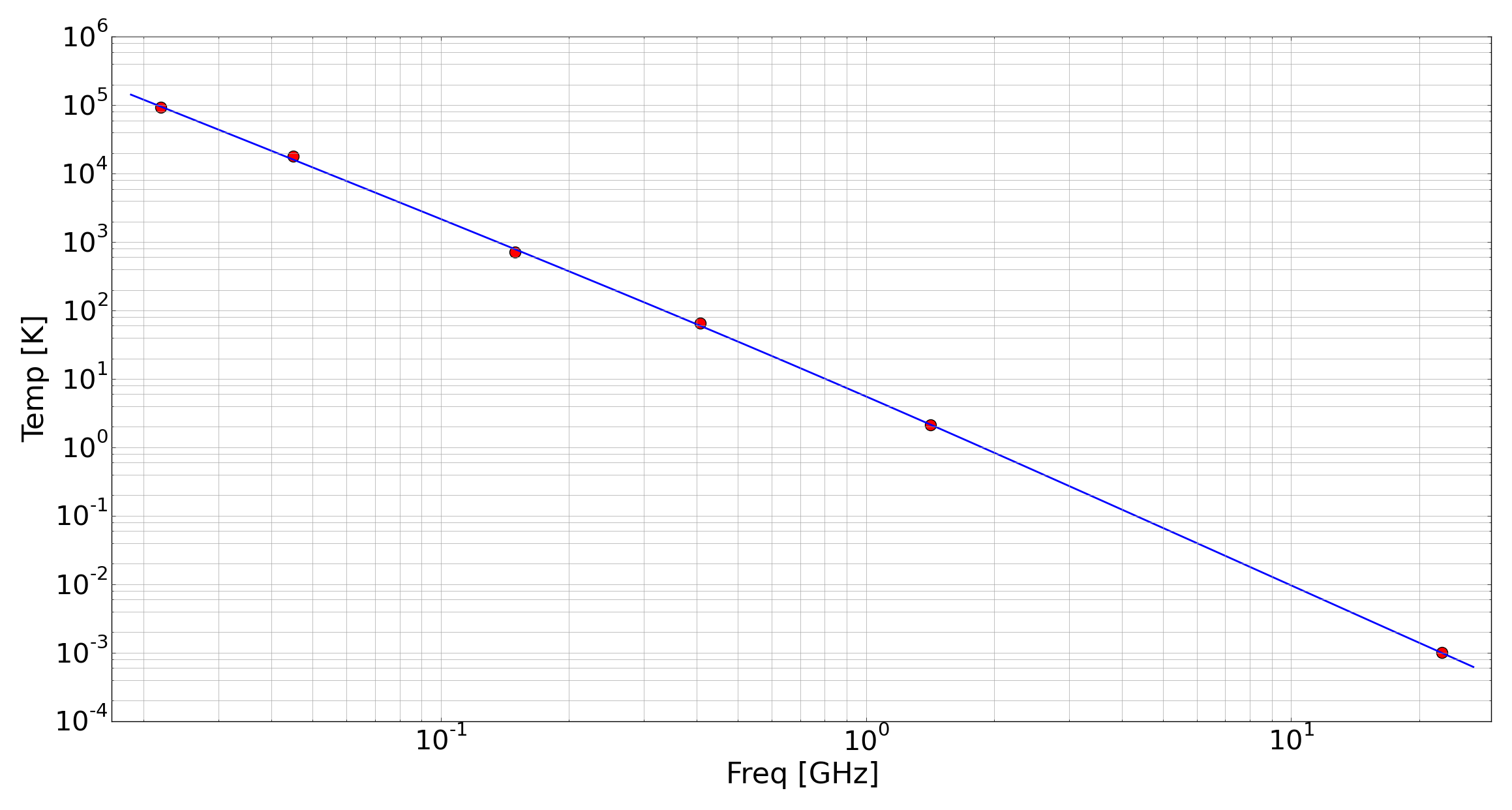}%
}

\subfloat[]{%
  \includegraphics[clip,width=\columnwidth]{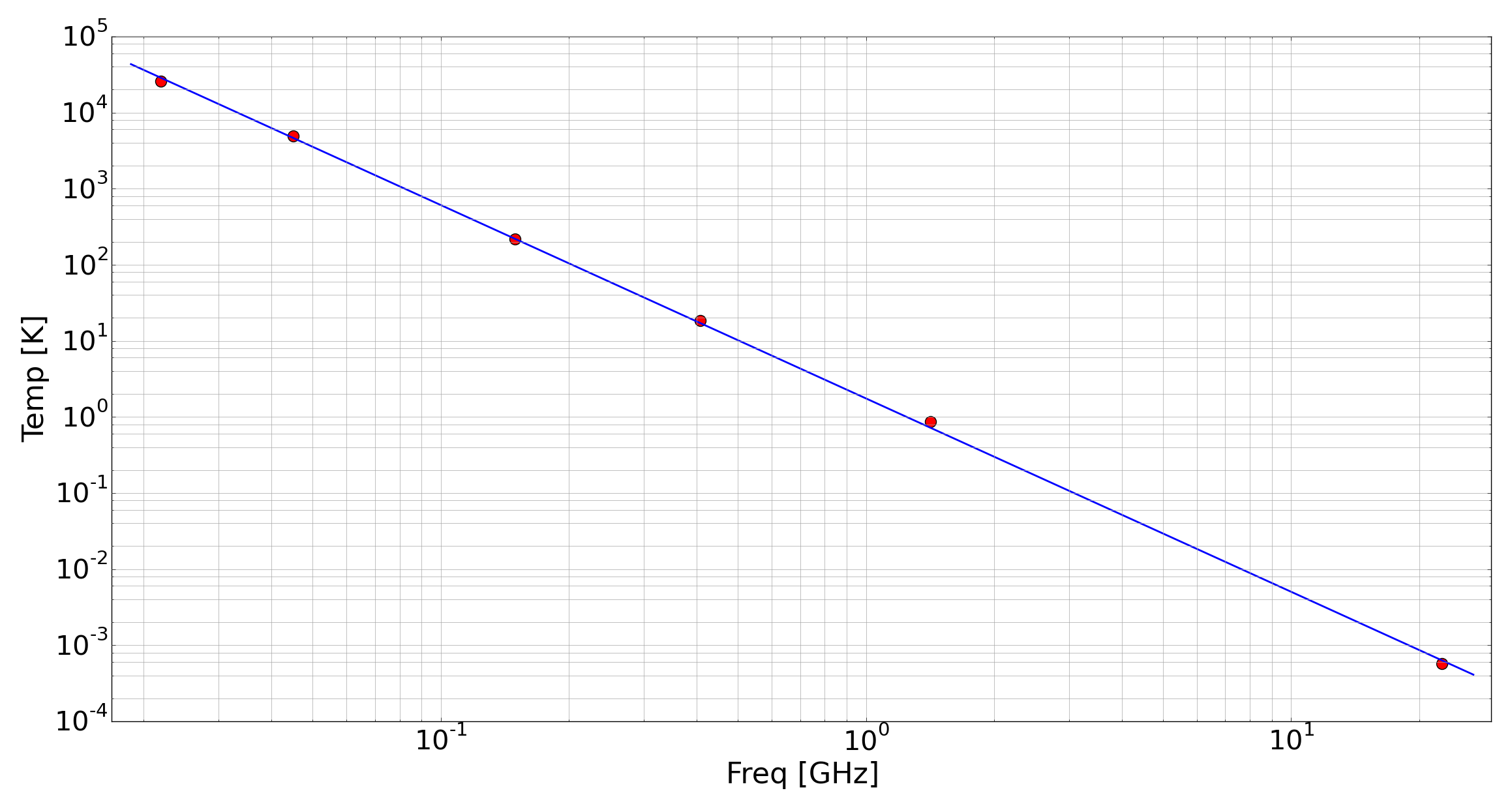}%
}
\end{minipage}
\begin{minipage}{0.5\textwidth}
\subfloat[]{%
  \includegraphics[clip,width=\columnwidth]{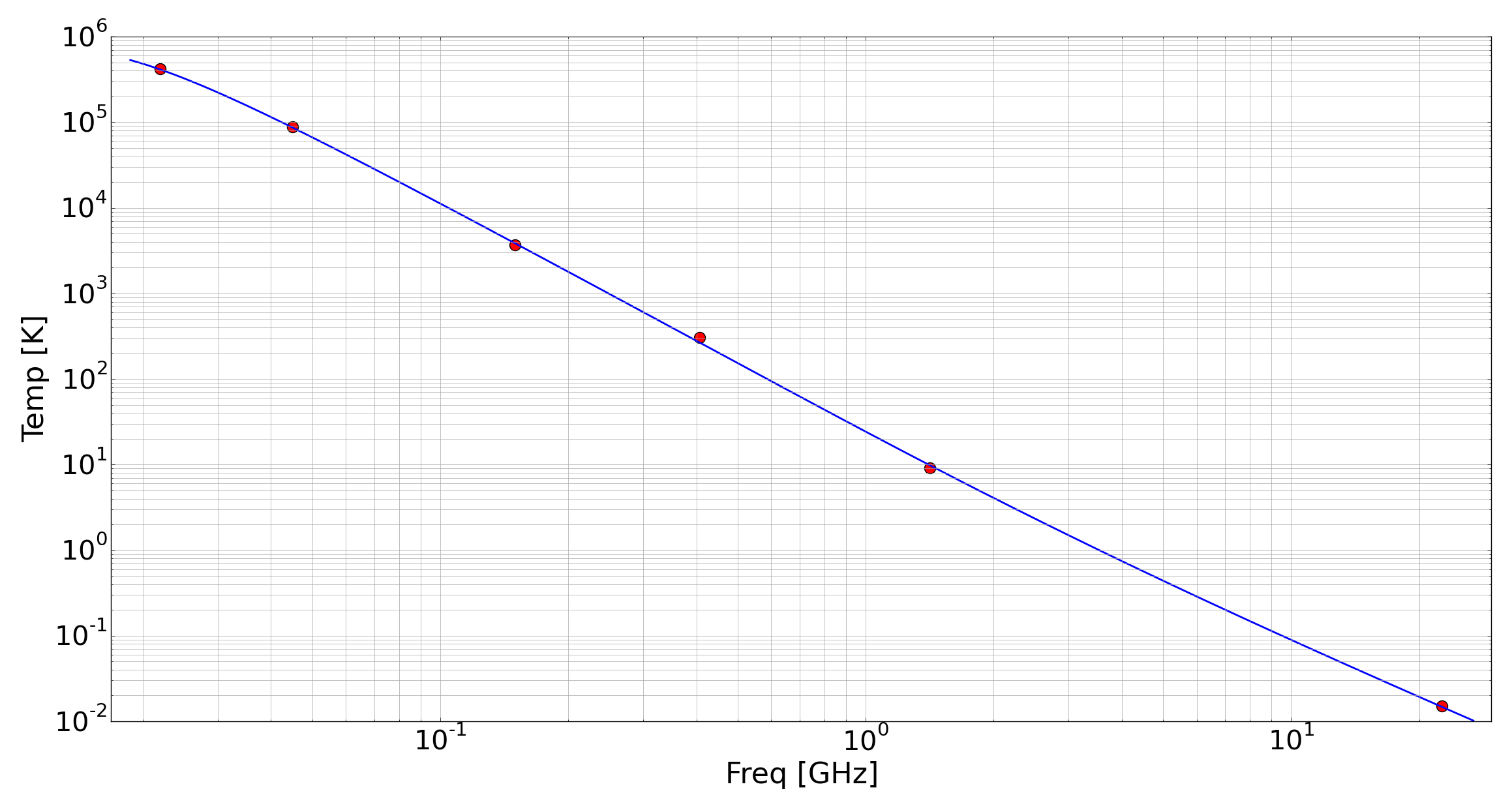}%
}

\subfloat[]{%
  \includegraphics[clip,width=\columnwidth]{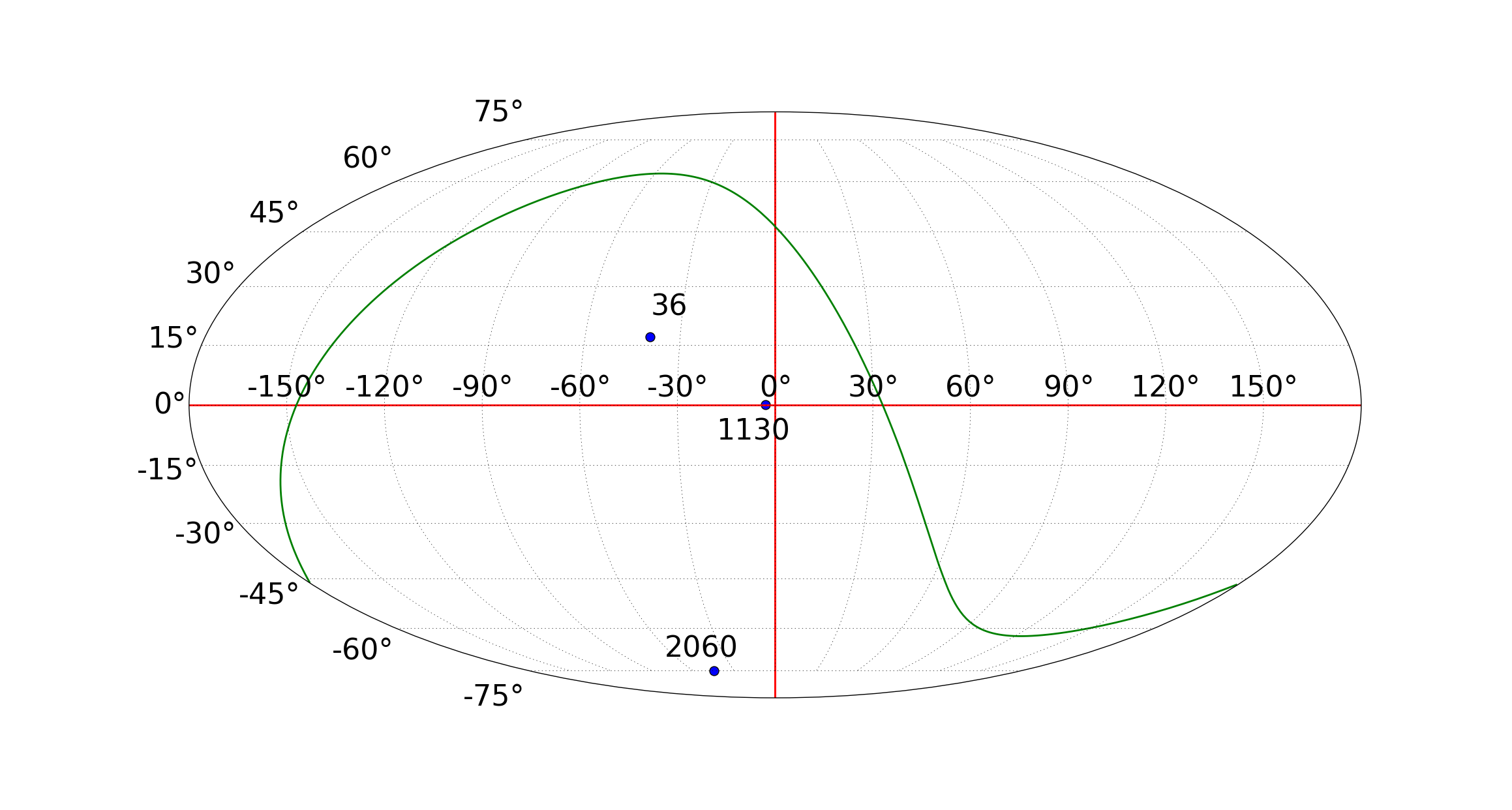}%
}

\end{minipage}

\caption{Data points (filled red circles) towards representative pixels with overlaid GMOSS generated spectra (solid blue lines).   Pixel positions selected for display contain (a) convex shape at pixel 36,  (b) concave shape at pixel 2060 and (c) complex shape at pixel 1130. The plots are in log-temperature versus log-frequency scale. Panel (d) shows the positions of these pixels on a Mollweide projection of the sky in Galactic coordinates, where the solid green line traces the ecliptic.}\label{fig:outpix}
\end{figure}

Shown in Figure~\ref{fig:outpix} are sample spectra derived from GMOSS (blue solid lines) towards three pixels that are representative of convex (pixel 36), concave (pixel 2060) and more complex (pixel 1130) spectral shapes. The measurement data from the six maps are also shown in the three panels using filled red circles.  Figure~\ref{fig:outpix} also shows in a panel the locations of these pixels as an image in Galactic coordinates. The pixels chosen lie towards different sky regions, where the spectral shapes are dominated by emission from different components.  As demonstrated by the fit solution to sample pixel 1130, the emission from the Galactic center region is, as expected, not only brighter than emission away from the plane, but the spectral structure too is more complex necessitating modeling with a broken power-law synchrotron emission, plus significant low-frequency turnover due to thermal absorption and excess free-free emission at the high-frequency end. This is to be expected from the variety of components and processes that are unique to the Galactic plane, particularly the Galactic center, including HII regions and supernova remnants to name a few.  Pixel 36 is in the vicinity of the central bulge, and the fit solution is that of a convex spectrum in which the emission is modeled in GMOSS as a broken power-law synchrotron spectrum, without significant thermal effects needed, which are mainly present towards the Galactic plane. The spectrum at pixel 2060 is in the vicinity of the Galactic pole and has a concave form. The spectrum towards this pixel is modeled by GMOSS as composite emission from flat and steep spectrum components. 

\begin{figure}[htp]
\begin{minipage}{0.5\textwidth}

\subfloat[]{%
  \includegraphics[clip,width=\columnwidth]{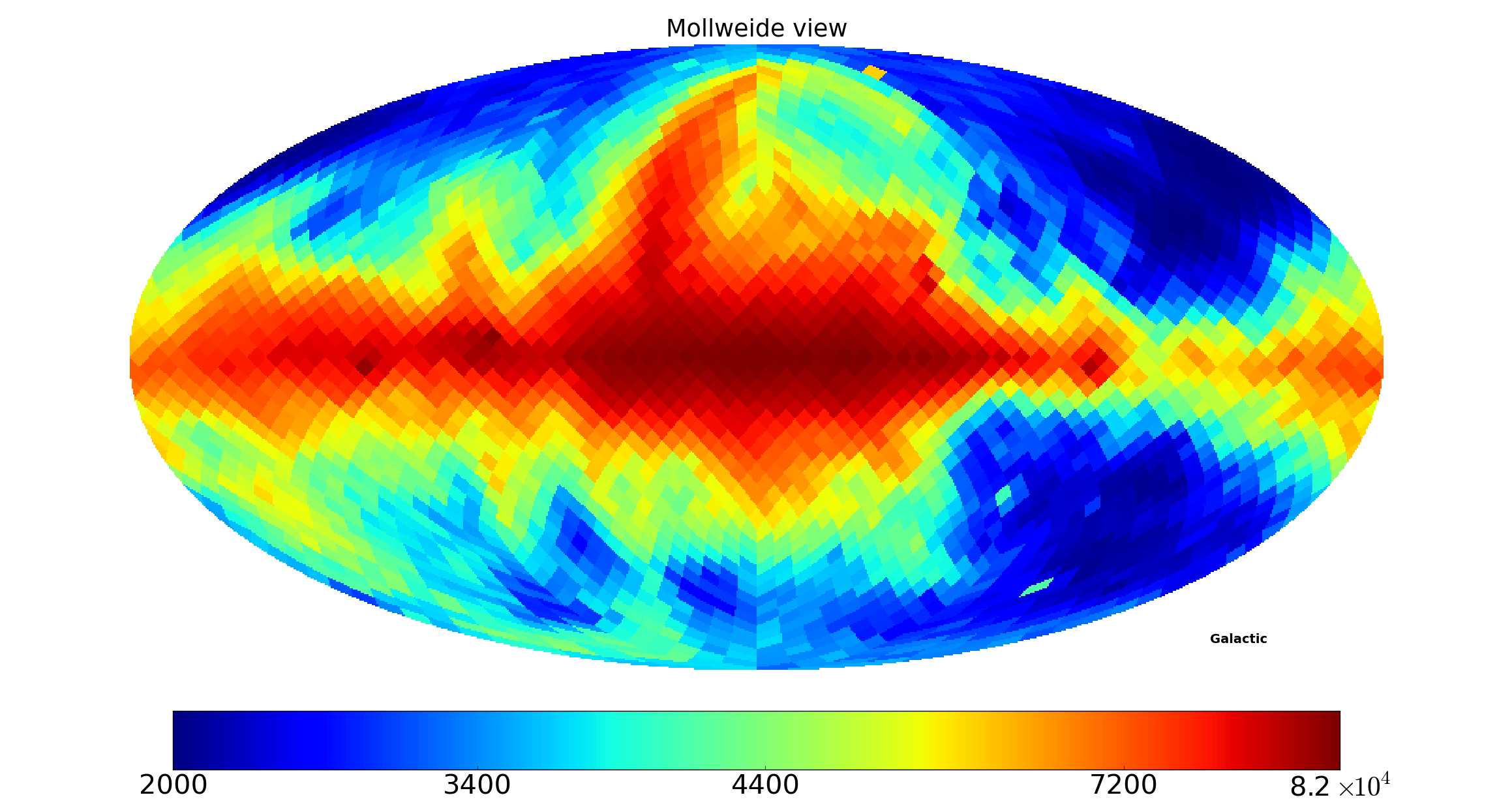}%
}
\end{minipage}
\begin{minipage}{0.5\textwidth}

\subfloat[]{%
  \includegraphics[clip,width=\columnwidth]{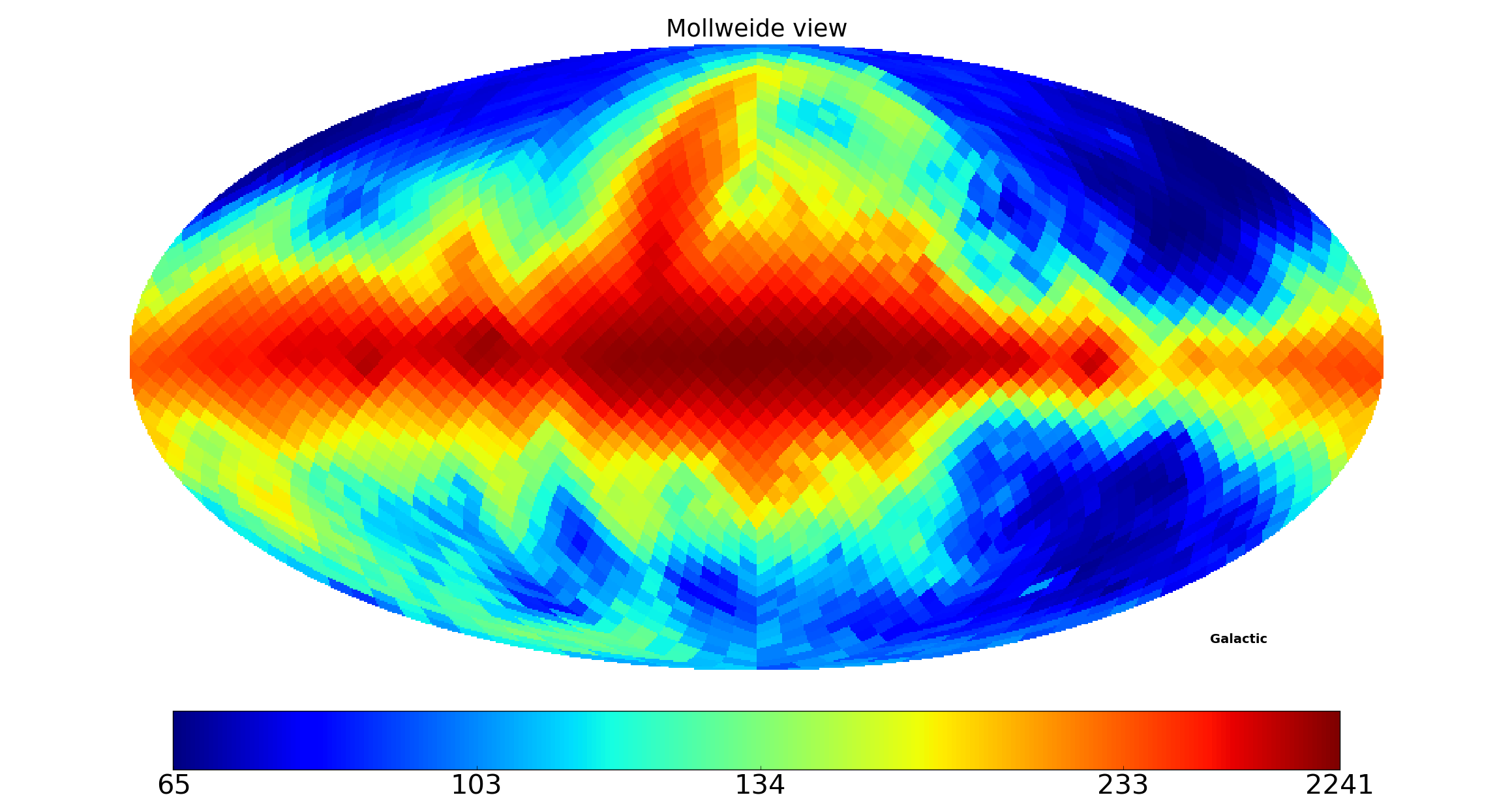}%
}
\end{minipage}
\caption{All-sky maps derived from GMS at (a) 50 MHz, (b) 200 MHz.  The maps are in units of Kelvin at a resolution of $5^{\circ}$ and in Galactic coordinates.}\label{fig:GMOSS_ASM}
\end{figure}

Two sample all-sky maps at 50 MHz and 200 MHz generated using GMOSS are shown in Figure~\ref{fig:GMOSS_ASM}. The Mollweide projection maps are in Galactic coordinates with 5$^\circ$ resolution. With this resolution the coarse Galactic features clearly arise in the maps, and the mean temperatures are higher in the lower frequency map as expected. These maps have a median deviation from their nearest input maps namely the ones at 45 MHz and 150 MHz by 25$\%$ and 50$\%$ respectively. 

\begin{figure}[htp]
\begin{minipage}{0.5\textwidth}
\subfloat[]{%
  \includegraphics[clip,width=\columnwidth,height=1.5in]{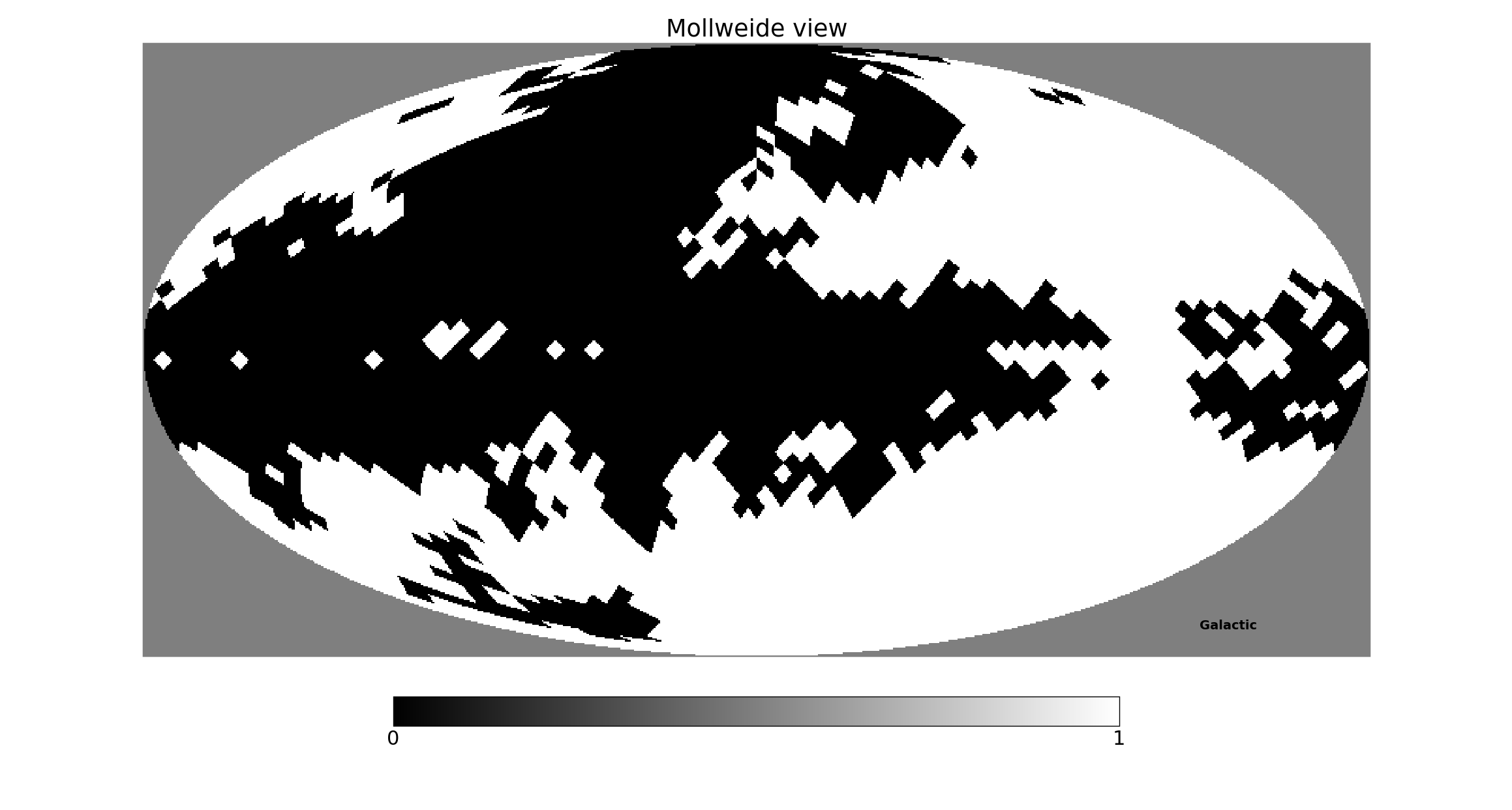}%
}

\subfloat[]{%
  \includegraphics[clip,width=\columnwidth,height=1.5in]{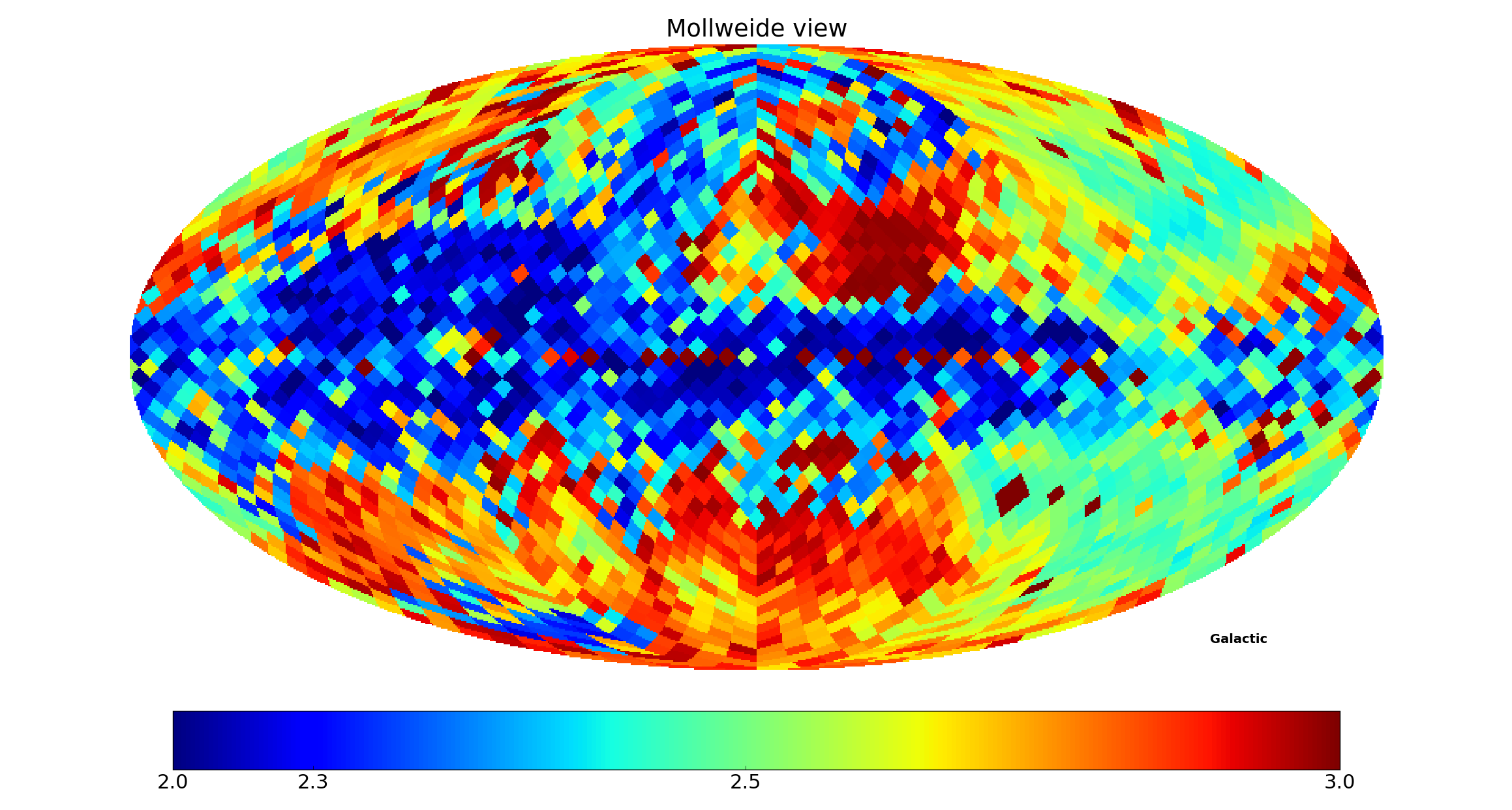}%
}

\subfloat[]{%
  \includegraphics[clip,width=\columnwidth,height=1.5in]{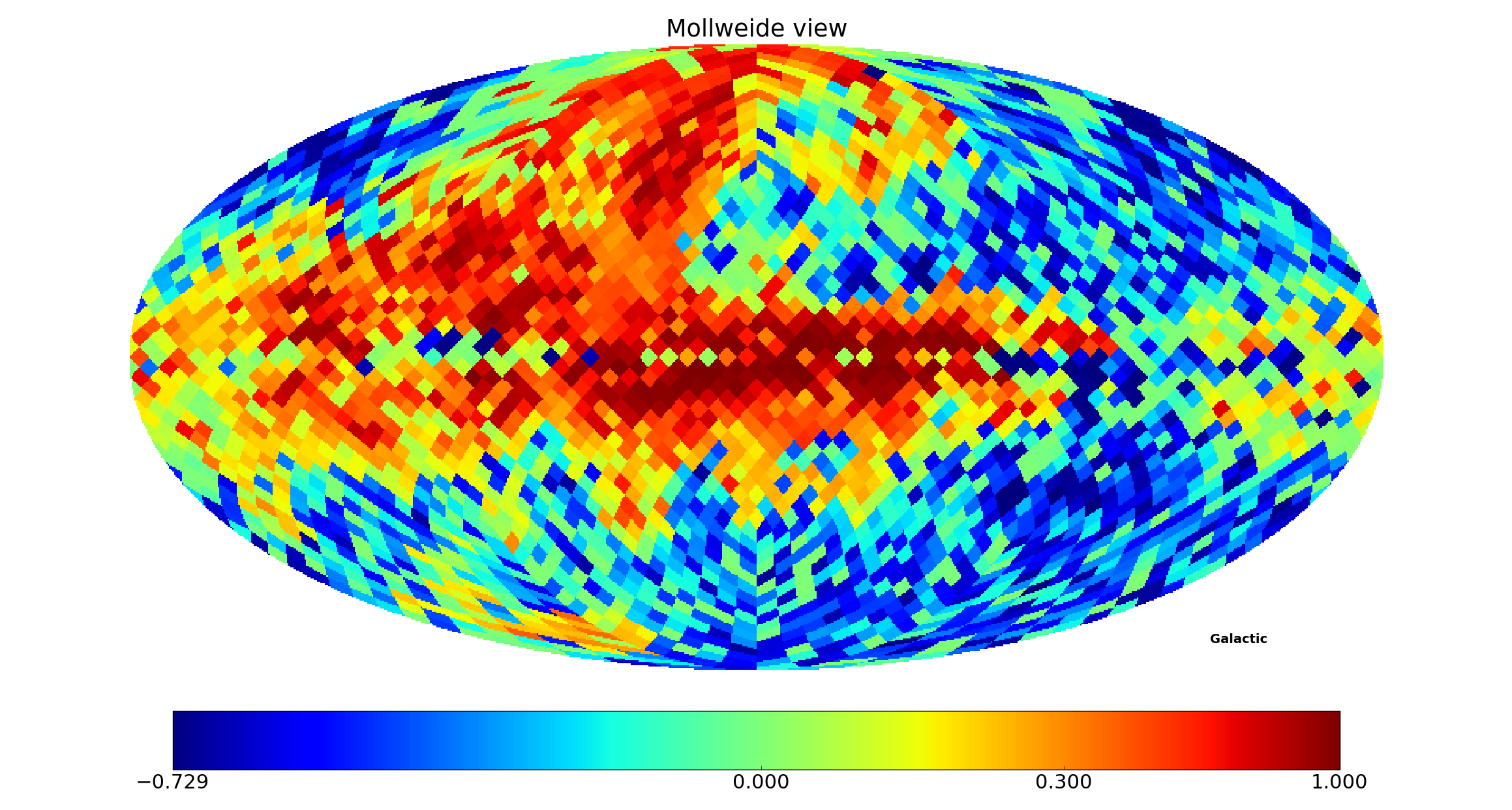}%
}

\subfloat[]{%
  \includegraphics[clip,width=\columnwidth,height=1.5in]{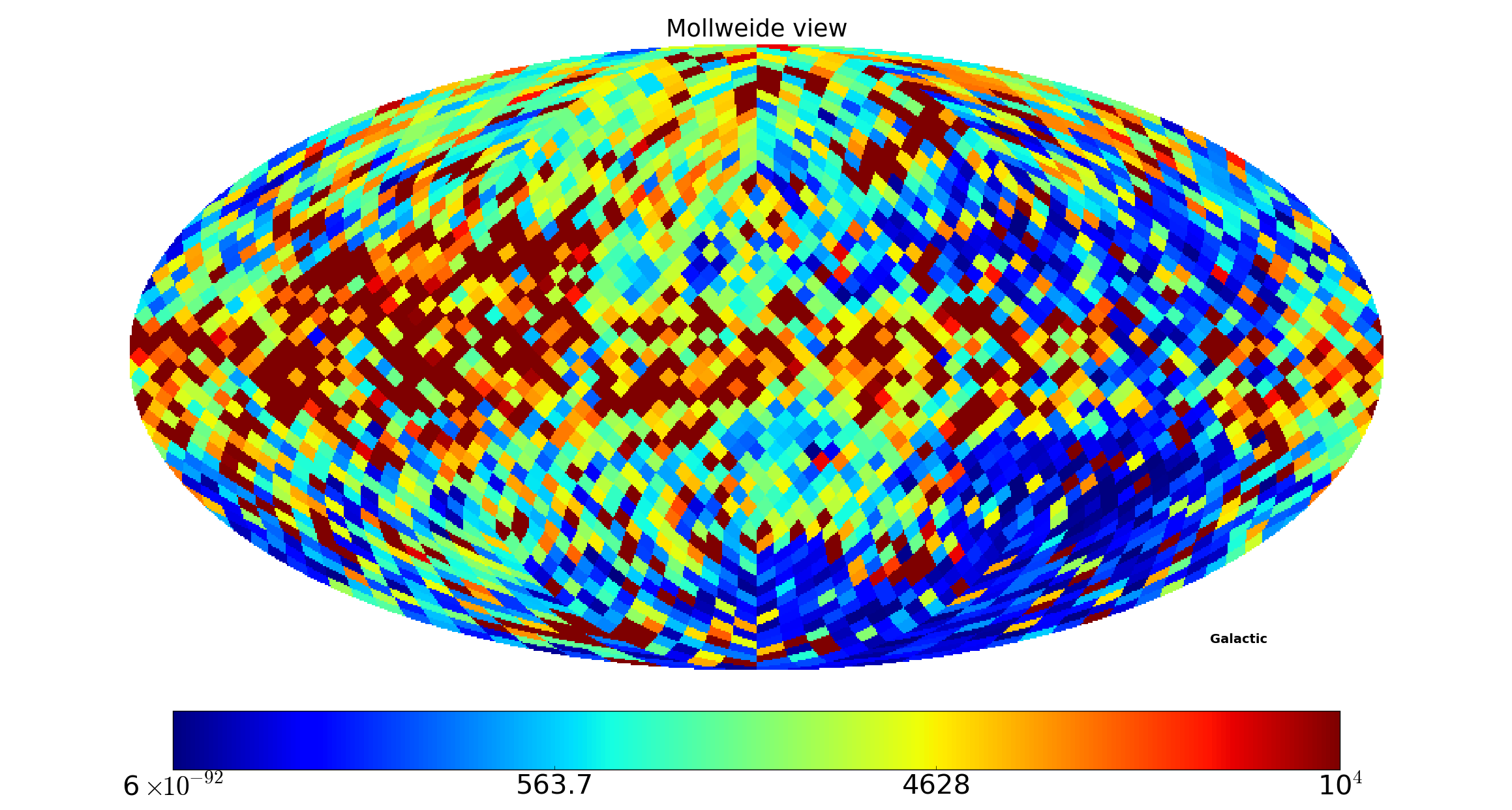}%
}
\end{minipage}
\begin{minipage}{0.5\textwidth}
\subfloat[]{%
  \includegraphics[clip,width=\columnwidth,height=1.5in]{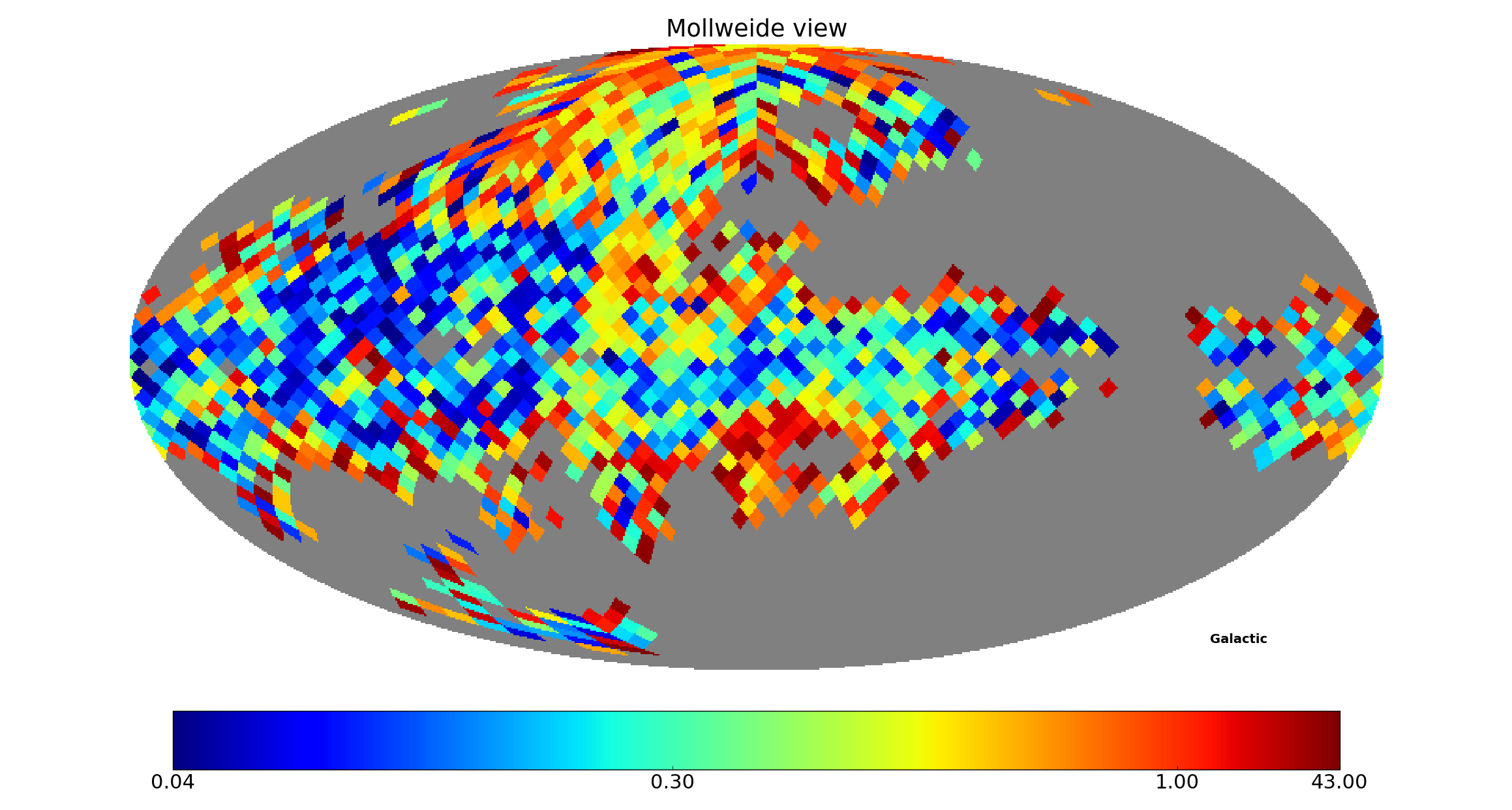}%
}\label{fig:nu_br}

\subfloat[]{%
  \includegraphics[clip,width=\columnwidth,height=1.5in]{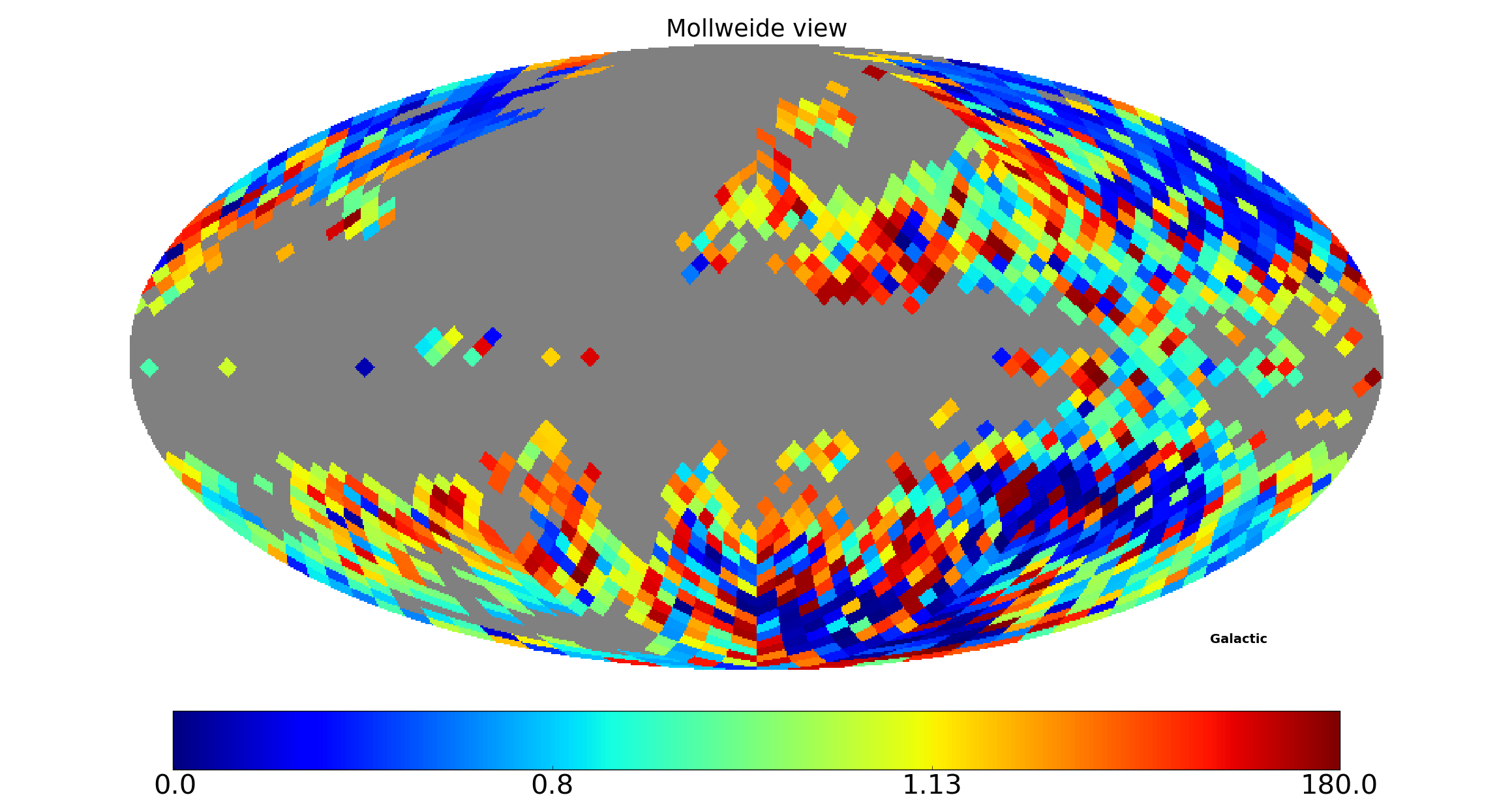}%
}\label{fig:norm2}

\subfloat[]{%
  \includegraphics[clip,width=\columnwidth,height=1.5in]{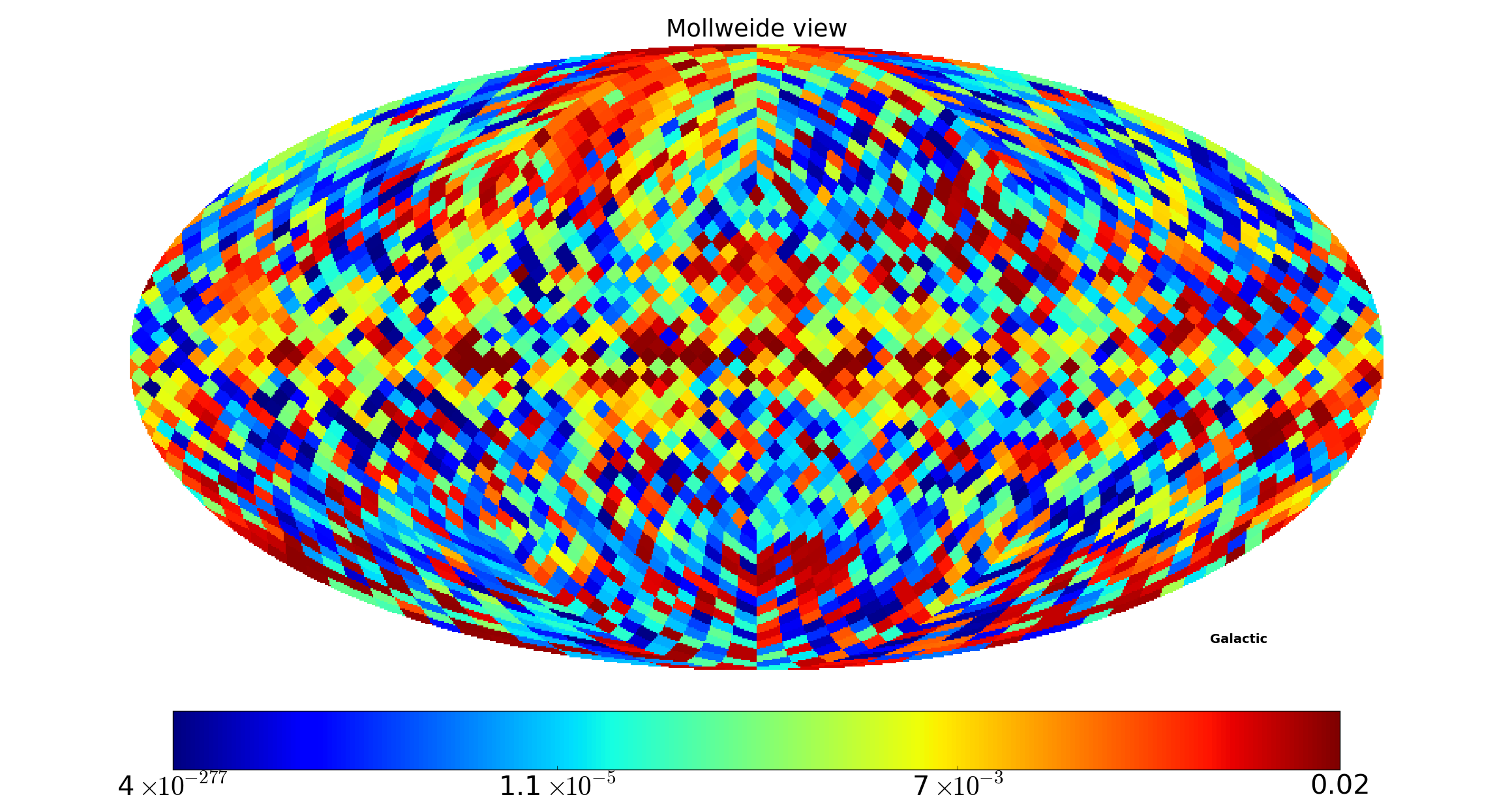}%
}

\subfloat[]{%
  \includegraphics[clip,width=\columnwidth,height=1.5in]{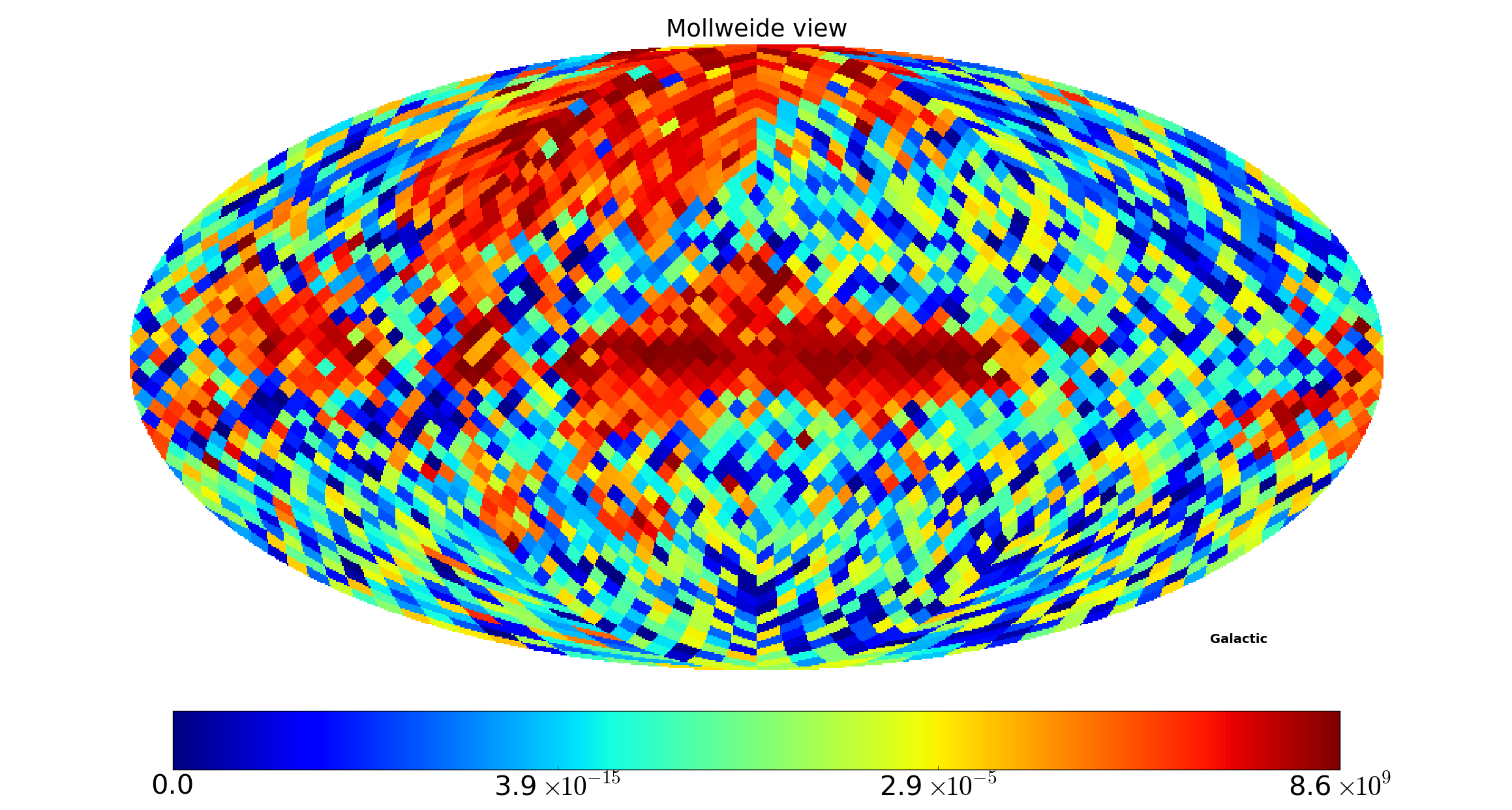}%
}
\end{minipage}
\caption{The sky distribution of the optimized values of parameters.  All maps are at a resolution of $5^{\circ}$ and in galactic coordinates. (a) Distribution of pixels with convex (black) and concave (white) spectra. The Galactic plane and north polar spur are clearly of distinct type suggesting different physics in the electron populations compared to the extragalactic sky (b) $\alpha_{\rm 1}$ (c) $\delta_{\rm \alpha}$ given as positive when $\alpha_{\rm 2}>\alpha_{\rm 1}$ and negative when $\alpha_{\rm 2}<\alpha_{\rm 1}$ such that $\alpha_{\rm 2} = \alpha_{\rm 1}+\delta_{\rm \alpha}$ (d) Temperature $T_{\rm e}$ (e) break frequency $\nu_{\rm br}$ for convex spectra in GHz (f) the additional normalization for the flat spectrum sources, $C_{\rm 2}$ for concave spectra (g) the frequency of thermal absorption turnover, $\nu_{\rm t}$ and (h) the parameter representing optically thin free-free emission, $I_{\rm x}$.}\label{fig:out_params}
\end{figure}

The panels in Figure~\ref{fig:out_params} show the distribution of optimized parameters across the 3072 sky pixels on Mollweide projections of the sky in Galactic coordinates. Panels showing the break frequency and secondary normalization parameters in convex and concave spectra respectively are mutually exclusive in modeling the sky spectrum in GMOSS. In each case, the pixels that have spectra of the other form are masked and are in grey. The median values of some of the optimized parameters are given in Table~\ref{table:med_params}.  
\begin{table}[htp]
\begin{center}
 \begin{tabular}{c c} 
 \hline\hline
 Parameter & Median value \\ [0.5ex] 
 \hline
 
 $\alpha_{\rm 1}$ & 2.50 \\ 
 
 $\alpha_{\rm 2}$ & 2.58  \\
 
 $\nu_{\rm br}$ & 0.36~{\rm GHz}\\
 
 $I_{\rm x}$ & $8.39\times10^{\rm -10}$  \\
 
 $T_{\rm e}$ & 2060~{\rm K} \\ 
 
 $\nu_{\rm t}$ & 0.3~{\rm MHz} \\ [1ex] 
  \hline
 \end{tabular}
\end{center}
\caption{Median values of optimized parameters over all 3072 pixels that describe the sky models.}\label{table:med_params} 
\end{table}

Figure~\ref{fig:out_params}(a) shows the relative distributions of pixels that have convex and concave spectra and thus entail two different models for the relativistic electrons.  It is interesting to note that these pixels employing different models are distributed towards distinct regions of the sky. The Galactic plane and the north polar spur are distinct and require convex spectra employing a break in the power-law synchrotron emission, with the spectral fit towards pixel 36 in Figure~\ref{fig:outpix} serving as a representative example. In contrast, the radio spectrum in regions off the Galactic plane and the spur are concave and hence required a modeling as a composite of flat and steep spectrum components; the spectral fit towards pixel 2060 in Figure~\ref{fig:outpix} is a representative example of this type.

The above difference is likely a reflection of the difference in the origin of the relativistic electron populations.  In the Galaxy the electrons are believed to be created in shock accelerations associated with supernovae and these then diffuse and migrate off the plane acquiring a break in the electron energy distribution because of aging and loss mechanisms \citep[][]{Lisenfeld2000}.  On the other hand, the extragalactic radio emission is dominated at low frequencies by powerful radio galaxies in which the acceleration is in hot spots at the ends of relativistic jets and perhaps in-situ re-acceleration in cocoons. At high frequencies, the dominant emission is from flat spectrum cores of active galactic nuclei \citep[see, for example][]{Miley1980}. This consistency between GMOSS results and expectations lends confidence to the modeling presented here.

Figure~\ref{fig:out_params}(b) gives the distribution of temperature spectral index $\alpha_{\rm 1}$ of the low-frequency synchrotron emission across the sky. The parameter $\delta_{\rm \alpha}$ that represents the change in spectral index towards high frequencies is shown in Figure~\ref{fig:out_params}(c). Unsurprisingly pixels that have positive values of $\delta_{\rm \alpha}$ are those that have convex spectra and are represented in black in Figure~\ref{fig:out_params}(a) and those with negative values correspond to pixels that have spectra of the concave shape, given by white pixels in Figure~\ref{fig:out_params}(a). 

The temperature of the thermal medium that models the absorption at low frequencies is shown in Figure~\ref{fig:out_params}(d) and the frequency of the thermal absorption turnover is in Figure~\ref{fig:out_params}(g). Pixels with the highest temperatures and the highest turnover frequencies lie mostly along the Galactic plane and towards the north polar spur.  Pixels in blue in Figure~\ref{fig:out_params}(g) have very low values, indicating that the extragalactic sky does not require any significant thermal absorption in the physical modeling. The median thermal absorption turnover frequency, $\nu_{\rm t}$, in the GMOSS modeling is 0.3~MHz, which is consistent with the estimates in literature \citep[][]{Novaco1978,Cane1979} that this value is in the ball park of 1~MHz.  The median value of the GMOSS parameter $T_{\rm e}$  for the ionized medium (WIM) that models this turnover is 2060~K, which is in the range of electron temperatures estimated for the interstellar medium:  from observations of radio recombination lines at 328, 75 and 34.5~MHz \citet{kantharia2001} estimate the extended low density warm ionized medium (ELDWIM) responsible for Carbon recombination lines to have $T_{\rm e}$ as low as 30--300~K, and based on 1.4~GHz recombination lines \citet{heiles1996} estimate the ELDWIM to have $T_{\rm e}$ of 7000~K.

Figures~\ref{fig:out_params}(e) and \ref{fig:out_params}(f) represent two mutually exclusive parameters of the physical models in GMOSS, namely, the break frequency $\nu_{\rm br}$ for pixels that have convex spectra and a secondary normalization $C_{\rm 2}$ for the flat-spectrum emission component in the case of pixels that have concave spectra. The median value of $\nu_{\rm br}$ is $360$ MHz (see Table \ref{table:med_params}). This is consistent with a pure diffusion model of Galactic cosmic ray electron propagation in \cite{Strong2011}, for a break in the electron injection spectrum of 4.0 GeV, if the magnetic field is 1.4~$\mu$G. 

Lastly, Figure~\ref{fig:out_params}(h) shows the parameter $I_{\rm x}$, which represents optically-thin free-free emission that dominates at high frequencies. This parameter clearly arises from a component that lies along the north polar spur and a region of the Galactic plane with a relatively smaller scale height. The pixels with spectra showing spectral steepening as well as significant optically thin free-free emission would most certainly have a spectral shape that is complex, due to the convex spectrum at low frequencies followed by the upturn at high frequencies where the excess emission dominates; the spectrum towards pixel 1130, as shown in Figure~\ref{fig:outpix}, is an example of this type.

\begin{figure}[htp]
\centering
\subfloat[]{%
  \includegraphics[clip,width=\columnwidth]{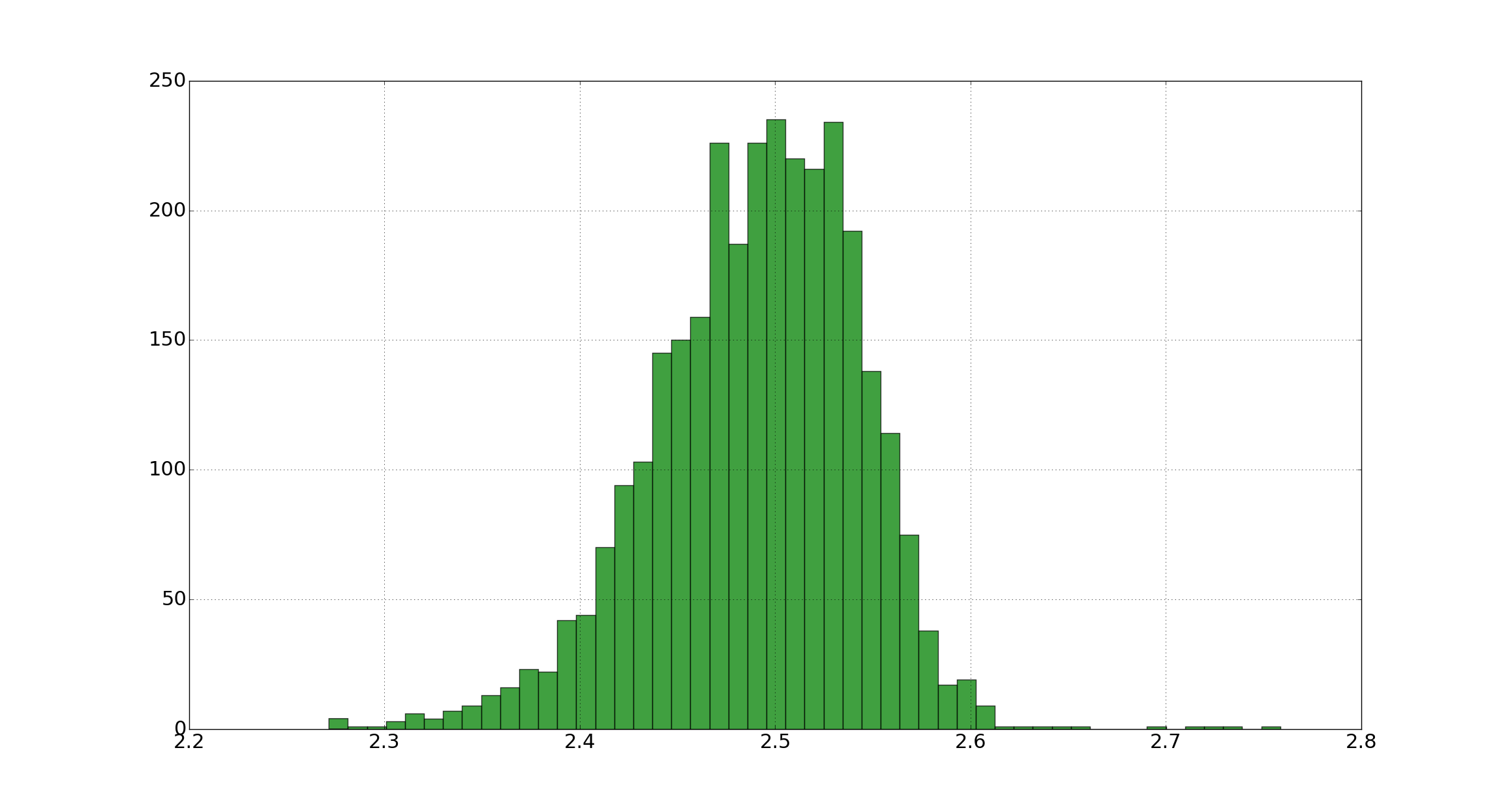}%
}

\subfloat[]{%
  \includegraphics[clip,width=\columnwidth]{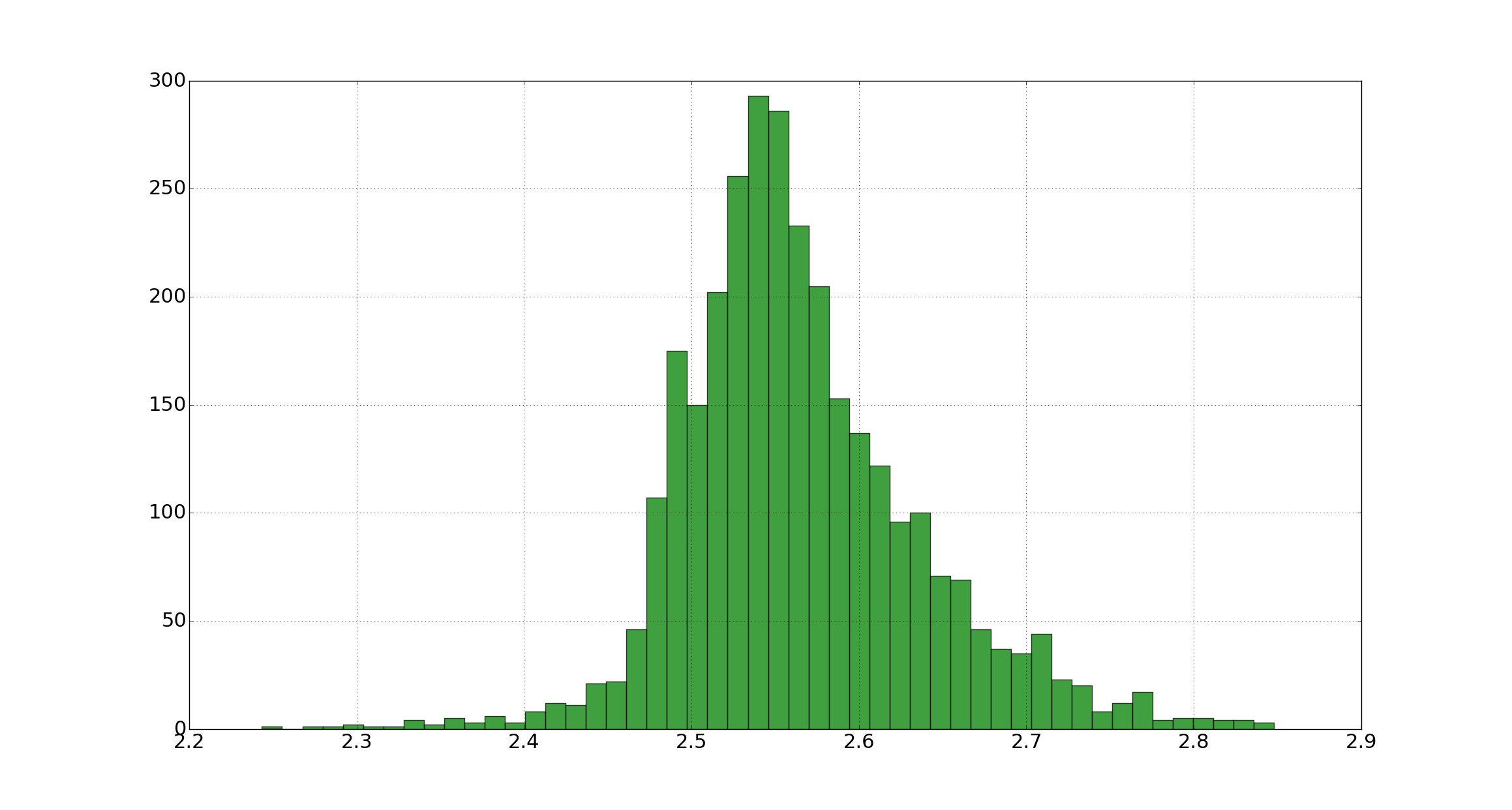}%
}
\caption{Histogram of 2-point spectral indices derived from GMOSS at all 3072 pixels between (a) 50 MHz and 150 MHz (b) 400 and 1200 MHz.}
\label{fig:si_hist}
\end{figure}

Using GMOSS generated model spectra at each pixel, we have computed the 2-point spectral indices for the pair of frequencies 50 and 150 MHz, and separately using the pair 400 and 1200 MHz.  Histograms of these 2-point spectral indices are shown in Figure \ref{fig:si_hist}.  The computed spectral indices have a mean value of 2.49 at the lower set of frequencies, where the spectral indices are distributed over the range 2.27 to 2.76.  Within the errors, this is same as the median value of 2.50 for the $\alpha_{\rm 1}$ model parameter, listed in Table~\ref{table:med_params}, that represents the spectral index of the synchrotron component below any break.  This low-frequency spectral index is also consistent with observations of \citet[]{Rogers2008}.   At relatively higher frequencies, the 2-point spectral index between 400 and 1200~MHz is 2.56 and in this band the range is over 2.24--2.85.  The median value of the spectral index parameter $\alpha_{\rm 2}$ above any break is 2.58 and in the same ballpark as the median 2-point spectral index.  The high frequency spectral index is again in agreement with measurements of the absolute radio sky at cm wavelengths \citep[see, for example,][]{Kogut2011}  .

\section{Summary}
\label{sec:conclusion}

We have presented a physically-motivated sky model of the low-frequency radio sky from 22~MHz to 23~GHz.   Spectra over this frequency range are presented over 3072 pixels covering the sky in R4 HEALPix pixels; the resolution is $5^{\circ}$.  GMOSS models the spectrum at each pixel primarily as optically thin synchrotron emission, either adopting a broken power-law form electron distribution or a composite of steep and flat spectrum components.  Additionally, optically-thin thermal emission is included to correct for any deficit at the high frequency end and thermal absorption is added as a foreground screen to account for any low-frequency flattening.  The 7-parameter model is fit to six all-sky images at 0.022, 0.045, 0.150, 0.408, 1.420 and 23~GHz to derive the GMOSS model spectra.  The fractional differences between GMOSS model spectra and the input data has a median value of $6\%$. This is in keeping with the systematic calibration errors in the input maps that range from 1--20~\%. Furthermore, derived physical parameters of the model  are in reasonable agreement with expected values, providing confidence in the physical model.

With a resolution of $5^{\circ}$, which is much finer than typical antenna beams used in experiments attempting to detect the global signature of EoR, the model can be convolved with appropriate antenna beams in simulations of these experiments to generate mock sky spectra.  GMOSS provides an expectation of the EoR foreground to help arrive at appropriate component separation strategies without any inherent assumptions of smoothness in the foreground spectrum. It may be noted here that GMOSS is intended to represent the radio continuum emission, and the physical processes included are those relevant to radio continuum in the MHz to GHz frequency range. This is useful to model the foreground contamination of wideband cosmological spectral signals or distortions of the CMB spectrum in these wavelengths. Radio recombination lines (RRLs) from the Galactic thermal component would be expected to add sharp spectral features, and hence be distinguishable from the wideband spectral signals such as those expected from recombination and reionization epochs and the Dark ages and Cosmic Dawn in between. Hence RRLs are not included in GMOSS. 

	Also, it may be noted that the all-sky radio maps used as inputs to GMOSS -- and indeed also all-sky maps that may be available in the foreseeable future -- do not have the accuracy anywhere near the level of the cosmological signals mentioned above. Including more images at intermediate frequencies in the coming years, as new images become available, will most certainly improve the GMOSS model.  If the fits yield residuals outside the error bars of the images, then more spectral complexity would necessarily have to be added to the model.  However, allowing for multiple breaks (which are indeed observed in sources where multiple cooling processes are presumed occurring) and/or multiple spectral components would most likely satisfy improved data, without requiring additional radiative processes.  It is not impossible that such inclusions might result in increased complexity of the synthetic spectra and hence prove to be of greater confusion to the detection of cosmological signals.

A study of the implications of GMOSS predicted sky spectra for EoR signal detection and signal extraction strategies are presented in a subsequent paper (Sathyanarayana Rao et al., in preparation). This subsequent paper presents among other things a comparison of GMOSS with existing polynomial models for foregrounds as well as a discussion on regions of the sky best suited for EoR signal detection experiments.   

However, since GMOSS is not tailored to EoR science, but is based entirely on plausible physics that produces the radio sky spectrum, it may be used for any problem that requires simulating the radio sky spectrum. Simulation studies of the foreground contamination in cm-wavelength detection experiments of signals arising from the epoch of recombination is one such problem where GMOSS can be applied. GMOSS is being made publicly available at {\tt www.rri.res.in/DISTORTION/} and will be updated as other maps and processes are included. The code used to generate the model is flexible and can include more maps, including those with partial sky coverage. 

\section{Acknowledgements}
We acknowledge the use of the Legacy Archive for Microwave Background Data Analysis (LAMBDA), part of the High Energy Astrophysics Science Archive Center (HEASARC). HEASARC/LAMBDA is a service of the Astrophysics Science Division at the NASA Goddard Space Flight Center. Some of the results in this paper have been derived using the HEALPix \citep{Gorski2005} package. MSR would like to thank K.~S.~Dwarakanath for useful discussions on Galactic radio emission. JC is supported by the Royal Society as a Royal Society University Research Fellow at the University of Manchester, U.K.


\end{document}